\documentclass{aa} 

\usepackage{graphicx}
\usepackage{subfig}
\usepackage{enumitem}
\usepackage{txfonts}
\usepackage{hyperref}
\usepackage{mathtools}

\def\frad{$f_{rad}$}
\begin{document} 

\title{Speed limits for radiation driven SMBH winds}
\author{
A. Luminari\inst{1,2},
F. Nicastro\inst{2,3},
M. Elvis\inst{3},
E. Piconcelli\inst{2},
F. Tombesi\inst{1,2,4,5},
L. Zappacosta\inst{2},
F. Fiore\inst{6},
}
\institute{
Department of Physics, University of Rome ``Tor Vergata'', Via della Ricerca Scientifica 1, I-00133 Rome, Italy \\ \email{\href{mailto:alfredo.luminari@inaf.it}{alfredo.luminari@inaf.it}}
\and
INAF - Osservatorio Astronomico di Roma, Via Frascati 33, I–00078 Monte Porzio Catone, Italy
\and
Harvard-Smithsonian Center for Astrophysics, 60 Garden Street, Cambridge, MA 02138, USA
\and
Department of Astronomy, University of Maryland, College Park, MD 20742, USA
\and
NASA - Goddard Space Flight Center, Code 662, Greenbelt, MD 20771, USA
\and
INAF - Osservatorio Astronomico di Trieste, Via G. B. Tiepolo 11,I–34143 Trieste, Italy}
\date{Received xxyyzz; accepted xxyyzz}

\abstract
{Ultra Fast Outflows (UFOs) have become an established feature in X-ray spectra of Active Galactic Nuclei (AGN). According to the standard picture, they are launched at accretion disc scales with relativistic velocities, up to 0.3-0.4 times the speed of light. Their high kinetic power is enough to induce an efficient feedback on galactic-scale, possibly contributing to the co-evolution between the central supermassive black hole (SMBH) and the host galaxy. It is therefore of paramount importance to fully understand the UFO physics, and in particular the forces driving their acceleration and the relation with the accretion flow from which they originate.}
{In this paper, we investigate the impact of special relativity effects on the radiative pressure exerted onto the outflow. The radiation received by the wind decreases for increasing outflow velocity $v$, implying that the standard Eddington limit argument has to be corrected according to $v$. Due to the limited ability of the radiation to counteract the black hole gravitational attraction, we expect to find lower typical velocities with respect to the non-relativistic scenario.}
{We integrate the relativistic-corrected outflow equation of motion for a realistic set of starting conditions. We concentrate on a range of ionisations, column densities and launching radii consistent with those typically estimated for UFOs. We explore a one-dimensional, spherical geometry and a three-dimensional setting with a rotating thin accretion disc. }
{We find that the inclusion of special relativity effects leads to sizeable differences in the wind dynamics and that $v$ is reduced up to 50 \% with respect to the non-relativistic treatment. We compare our results with a sample of UFO from the literature, and we find that the relativistic-corrected velocities are systematically lower than the reported ones, indicating the need for an additional mechanism, such as magnetic driving, to explain the highest velocity components. We note that these conclusions, derived for AGN winds, have a general applicability.}{}


\titlerunning{Speed limits for radiation driven SMBH winds}
\authorrunning{A. Luminari et al.}
\maketitle

\section{Introduction}
Fast outflows, and in particular Ultra Fast Outflows (UFOs), are routinely observed in Active Galactic Nuclei (AGN) as blueshifted absorption and emission features imprinted on the X-ray spectrum, with velocities ranging from $\sim 0.03$ to $0.4-0.5$ times the speed of light $c$ (\citealp{T11,T15,N15,F17,P17}). They are launched at accretion disc scales from the central supermassive black hole (SMBH) and may display a kinetic power as high as 20-40 \% of the bolometric luminosity of the AGN \citep{Feruglio15,N15,N18,La20}, which is more than enough to trigger a massive feedback in the host galaxy, according to theoretical models \citep{Dimatteo05,HE10,Ga11}.

Despite their crucial importance in the framework of the coevolution between the host galaxy and the SMBH \citep{KH13}, the physics of these outflows, and particularly their acceleration mechanism, still remains mostly unknown.
According to one of the most accepted scenarios, the gas is accelerated through the pressure of the radiation emitted in the vicinity of the central black hole, even though it is not yet fully clear up to which velocities this mechanism can be effective (\citealp{PK04, H14, Hagino15, KP15}).

Similarly, broad absorption lines (BALs) in the UV regime are observed in $\sim 10 -20 \%$ of optically selected quasars. Their velocities can be up to $0.3 c$ and are located at parsec-scales from the SMBH \citep{hamann18,bruni19}, thus representing another potential energy input for a galactic-scale feedback. As for the X-ray winds, radiative driving has been suggested as their main driver \citep{elvis00,matthews20}

In a recent paper, \citet{L20} discussed the importance of special relativity effects when the wind outflow velocity becomes mildly relativistic, $v \gtrsim 0.05 c$. In fact, due to the space-time transformation, the amount of radiative power received by the fast wind decreases with increasing $v$: with respect to a layer of gas at rest, the amount of radiation impinging on the wind is reduced of $\sim 30 \%$ for $v=0.1\ c$, and of $\sim 90 \%$ for $v=0.5\ c$. This implies that the classical derivation of the radiative pressure for a static gas is no longer valid for high velocity winds; accordingly, the radiative driving scenario has to be revised to incorporate these effects.

In this Paper we integrate the equation of motion for a wind launched at accretion disc scales in order to assess the impact of special relativity effects on radiative acceleration. We will mainly focus on X-ray winds; however, our results applies also for BAL winds.
In Sect. \ref{sect_1D} we present a simple, one-dimensional model of a wind illuminated by a luminosity corresponding to the Eddington value and we demonstrate that, owing to these effects, radiation alone is not able to counteract the gravitational attraction of the SMBH. In Sect. \ref{sect_2D} we present a three-dimensional scenario in which the gas is lifted from a geometrically thin accretion disk and it is radiatively accelerated. We refine this model in Sect. \ref{sect_frad} and Sect. \ref{n_steps} with a more detailed treatment of the wind opacity. We present the results in Sect. \ref{Results} and we discuss them in Sect. \ref{Discussion}. Finally, we summarise our results in Sect. \ref{conclusion}.

\section{One-dimensional, spherically symmetric wind model}
\label{sect_1D}
In order to have a glimpse on the importance of special relativity effects, we start from a simple toy model as follows. We assume that all the luminosity comes from a central point source and the gas has an initial velocity $v_0$ at a distance $r_0$ from the centre. We solve the equation of motion along the radial coordinate $r$.
According to the Euler momentum equation, the force exerted by a radiative pressure gradient $\nabla p$ on an infinitesimal wind element with density $\rho$ is:
\begin{equation}
\rho \frac{dv}{dt}=\nabla p -\frac{G M \rho}{r^2}
\label{main_eq_gen}
\end{equation}
where $\frac{dv}{dt}$ is the wind acceleration and $G,M$ are the gravitational constant and the SMBH mass, respectively. For the sake of simplicity, we assume for the moment that i) the wind is optically thin and ii) its opacity is dominated by the Thomson cross-section. We will relax these assumptions in the following. This way, we obtain an equation of motion for the wind:
\begin{equation}
\frac{dv}{dt}=\frac{L' k}{4\pi r^2 c}-\frac{G M}{r^2}
\label{main_eq}
\end{equation}
Where $L'$ is the central luminosity in the wind reference frame $K'$ and $k$ is the opacity of the wind, that we approximate as $k=\frac{\sigma_T}{m_p}$, where $\sigma_T, m_p$ are the Thomson cross-section and the proton mass, respectively (\citealp{RL}). In the definition of $L'$ we include special relativity effects as described in \citet{L20}, so that $L'=L\cdot \Psi$, where $L$ is the luminosity in the source frame $K$ and $\Psi \equiv \psi^4=\frac{1}{\gamma^4 (1+\beta cos(\theta))^4}$, where $\gamma$ is the Lorentz factor, $\beta=\frac{v}{c}$ and $\theta$ is the angle between the velocity of the gas and the incident luminosity $L$.

For a wind that is moving radially outward, $\theta=0 deg$, the luminosity can be written as $L'=L \frac{(1-\beta)^2}{(1+\beta)^2}$. We can rewrite Eq. \ref{main_eq} as:
\begin{equation}
\frac{dv}{dt}=L \frac{(1-\beta)^2}{(1+\beta)^2} \frac{\sigma_T}{4\pi r^2 c m_p}-\frac{G M}{r^2}
\label{red_eq}
\end{equation}
Where $r$ and $\beta$ are functions of $v$ itself. The complete set of equations, including initial conditions, can be written as:
\begin{subequations}
\begin{align}
	\frac{dv}{dt} & =L \frac{(1-\frac{v}{c})^2}{(1+\frac{v}{c})^2} \frac{\sigma_T}{4\pi r(t)^2 c m_p} -\frac{G M }{r^2}\\
	r(t) & = r_0+\int_{t_0}^{t_1} v\ dt \\
	r_0 & =r(t=t_0) \\
	v_0 & =v(t=t_0)
\end{align}
\label{system}
\end{subequations}
Where $t_0, t_1$ are the starting and ending time of the numerical integration, respectively.
Moreover, we assume that the launching velocity of the wind $v_0$ corresponds to the rotational velocity of an accretion disc, orbiting around the black hole with a Keplerian profile, at $r=r_0$, so that $v_0=\sqrt{\frac{GM}{r_0}}$. Albeit this choice of $v_0$ may seem arbitrary at this point, it will be useful to compare the results with those of the following Sections. We also note that the escape velocity at $r=r_0$ is equal to $\sqrt{\frac{2 GM}{r_0}}=\sqrt{2} v_0$ . We can rewrite Eqs. \ref{system} as:
\begin{subequations}
\begin{align}
	\frac{dv}{dt} & =\Big( \lambda_{Edd} \frac{(1-v)^2}{(1+v)^2} - 1 \Big) \frac{1}{r(t)^2}\\
	r(t) & = r_0+\int_{t_0}^{t_1} v\ dt \\
	r_0 & =r(t=t_0) \\
	v_0 & =\sqrt{\frac{1}{r_0}}
\end{align}
\label{system_red}
\end{subequations}
where $\lambda_{Edd} \equiv L/L_{Edd}$ is the luminosity in units of the Eddington luminosity $L_{Edd}=\frac{4 \pi GM m_p c}{\sigma_T}$, $r,t$ are in units of the gravitational radius and time, $r_G=\frac{GM}{c^2}, t_G=\frac{r_G}{c}$ respectively,  and $v$ is in units of $c$.
We span the interval between 5 and 500 $r_G$ for $r_0$, to encompass the typical launching radius of UFOs, which usually lies between $\sim 50$ and some hundreds $r_G$ (\citealp{T12, T13, N15, T15, La20}). We divide this interval in five logarithmically-spaced steps: $r_0 \in [5.0,15.8,50.0,158.1,500.0] r_G$.  As we will discuss in detail in Sect. \ref{n_steps}, we note that the assumption of a point source may be less accurate for launching radii smaller than $50.0 r_G$. Nonetheless, it is instructive to study the solutions down to the smallest radii to identify possible trends. We integrate the equation for $10^6 t_G$, after which the dynamics of the wind reaches a steady state and the velocity appears to be almost constant in all the cases. We note that $10^6 t_G$ corresponds to $\sim 1 (100) yrs$ for $M=10^7(10^9) M_{sun}$, while present-day X-ray observations have observation times smaller than a month. This will allow us to follow the wind dynamics for a sufficient time scale to compare with the observations for any value of $M$ inside the typical AGN range. We find that the wind evolution is best sampled by a logarithmic time grid, rather than by linear steps. We fix the number of time elements to $5 \cdot 10^6$ to obtain an optimal numerical accuracy. Using a higher resolution does not produce noticeable improvements in the solutions.
 
\begin{figure}
\centering
\includegraphics[width=\columnwidth]{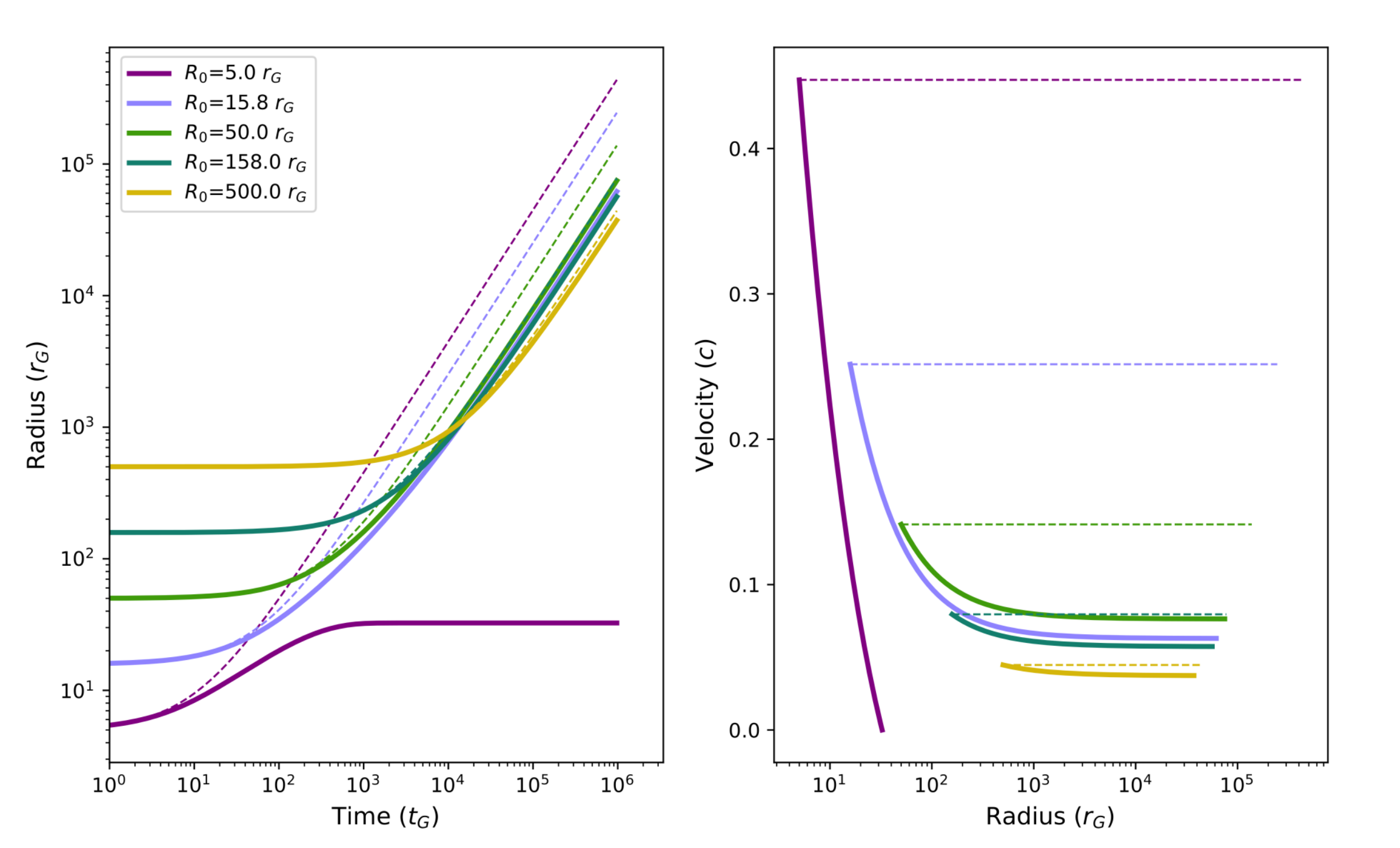}
\caption{Left: radial distance from the black hole $r(t)$ as a function of $t$ for $\lambda_{Edd}=1$ and $v_0=v_{rot}=\sqrt{\frac{GM}{r_0}}$; Right: radial velocity $v(t)$ as a function of $r(t)$. Solid lines indicate the trajectories in the relativistic framework (Eqs. \ref{system_red}), while dashed lines correspond to the classic (i.e., non-relativistic) ones.}
\label{kepler1}
\end{figure}

We show in Figure \ref{kepler1} the numerical result of Eqs. \ref{system_red} for $\lambda_{Edd}=1$. For comparison, we also show the classic analogue of Eq. \ref{system_red} , i.e., without the luminosity reduction factor $\Psi$ due to relativistic effects. Hereafter, we will indicate with solid(dashed) lines the values relative to the relativistic(classic) treatment, if not stated otherwise. Since we input a luminosity corresponding to the Eddington limit, $\lambda_{Edd}=1$, in the classic case the radiative pressure is able to counteract (by definition) the gravitational pull from the black hole. As a result, the acceleration of the gas is null and we obtain constant velocity solutions. Once the wind is launched with a given $v_0$, it escapes from the system with constant $v=v_0$, as can be seen in the right panel of Fig. \ref{kepler1}.

As expected, relativistic effects reduce radiation pressure, resulting in a deceleration of the wind under the gravitational pull of the SMBH. Indeed, it can be seen that the wind trajectories, especially the ones at smaller radii (i.e., closer to the black hole) undergo a significant velocity reduction. In the extreme case of $r_0=5 r_G$, the velocity drops to 0 . When $v=0$, relativistic effects vanish and radiation pressure is able, as in the classic case, to sustain the wind against the gravitational force, leading to a "stalling wind".
The highest final velocity is given by the case with $r_0=50.0 r_G$ and it is $\approx 0.1 c$, consistent with the typical observed UFO velocities (e.g., \citealp{T11, Goff15}).

\section{Axisymmetric wind launched from an accretion disk}
\label{sect_2D}
Let us now rewrite Eq. \ref{system_red} in the case of a wind launched from an accretion disc. We assume axisymmetry and a geometrically thin disc, such as in \citet{SS73}, orbiting with a Keplerian profile. We adopt a cylindrical coordinate system $(R,\phi,z)$. The set of equations is then:
\begin{subequations}
\begin{align}
	\frac{dv_R}{dt} & =\Big( \frac{\lambda_{Edd}}{\gamma^4(1+\beta cos\theta)^4} - 1 \Big) \frac{R}{r^3} + \frac{l^2}{R^3}\\
	\frac{dv_z}{dt} & =\Big( \frac{\lambda_{Edd}}{\gamma^4(1+\beta cos\theta)^4} - 1 \Big) \frac{z}{r^3}\\
	R & = R_0+\int_{t_0}^{t_1} v_R\ dt \\
	z & = \int_{t_0}^{t_1} v_z\ dt \\
    r & = \sqrt{R^2+z^2} \\
    \vec{v}_0 & =\vec{v}(t=t_0) = (0,v_{rot},v_{z,0}) \\
    \vec{r}_0 & = \vec{r}(t=t_0) = (R_0,0,0)
\end{align}
\label{2D_pre}
\label{2D}
\end{subequations}
where, as in Eqs. \ref{system_red}, $r,t$ are in units of $r_G,t_G$, $\theta$ is the angle between the incident luminosity and the direction of motion of the gas and we assume that the luminosity source is point-like. Hereafter, we indicate in bold the vectorial quantities. In the first two equations, the second term in the right-hand bracket corresponds to the gravitational attraction. $l$ is the specific angular momentum (angular momentum per unit mass), which is a conserved quantity during the motion. $\vec{r}_0, \vec{v}_0$ are the starting radius and velocity, respectively. We assume that, initially, the gas lies on the disk plane and the starting velocity lifts the gas above the disk, along the $z$ coordinate, with a velocity proportional to the disc rotational velocity $v_{rot}=\sqrt{\frac{1}{R_0}}$. The velocity along the $\phi$ coordinate is updated at each step to ensure the conservation of $l$. These initial conditions are rather general, and represents a good approximation of the radiatively-driven wind scenario (\citealp{PSK00,PK04}), as well as the magneto-hydrodynamic (MHD) scenario, in which the gas is lifted through magnetic field lines co-rotating with the disk (\citealp{BP82,CL94,F10,F14,Cui20}).

\begin{figure}
\centering
\includegraphics[width=\columnwidth]{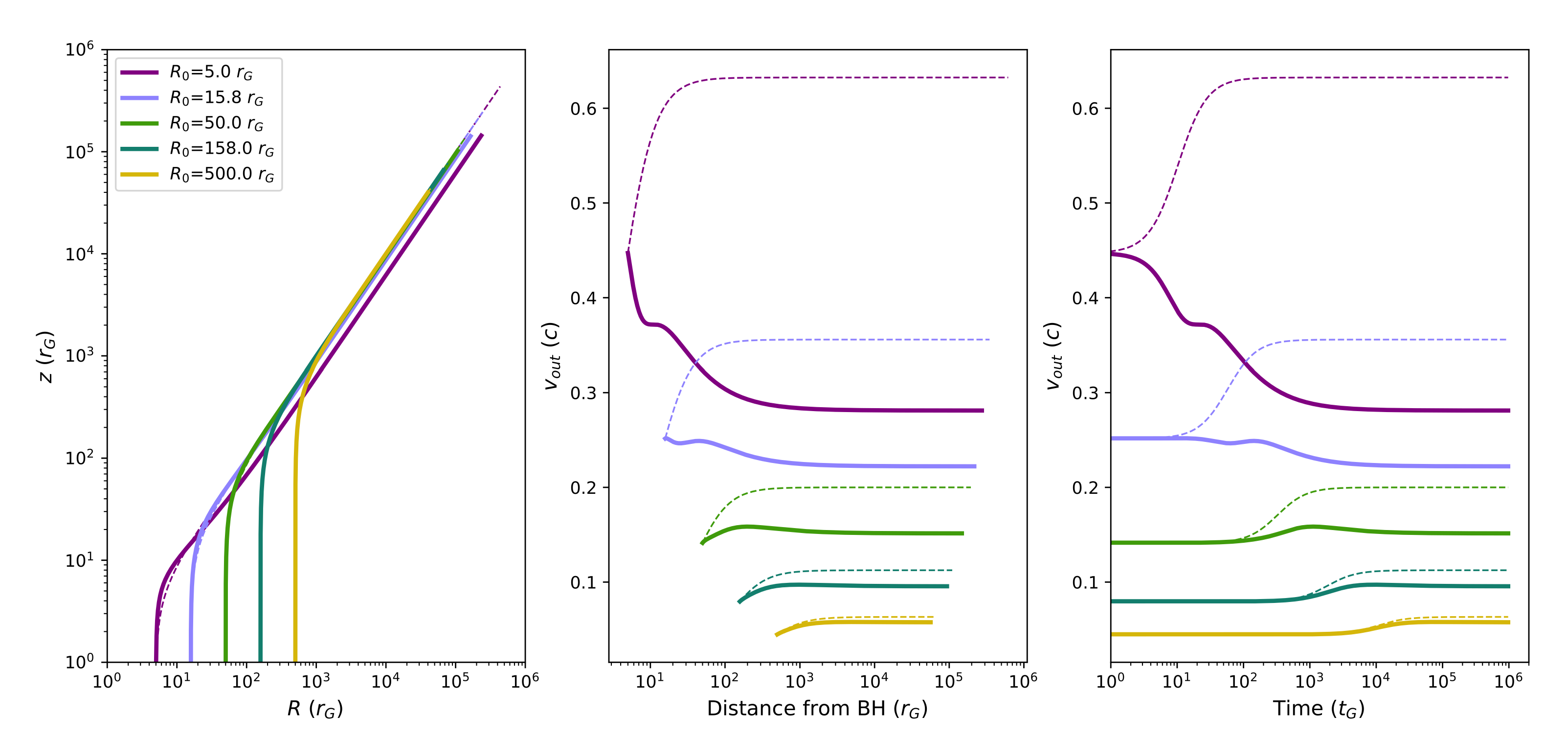}
\caption{Wind trajectories for $v_{z,0}=v_{rot}$ and $\lambda_{Edd}=1$ in axisymmetric geometry. From left to right: trajectories of the wind in the $z-R$ plane (x-axis corresponds to the radius and y-axis to the altitude from the disc plane); outflow velocity $v_{out}$ as a function of $r$ (where $r=\sqrt{R^2 + z^2}, v_{out} = \sqrt{v_R^2+v_z^2}$); $v_{out}$ as a function of $t$. Solid(dashed) lines refer to the relativistic(classic) treatment.}
\label{kepler-red}
\end{figure}

Fig. \ref{kepler-red} shows the solutions of Eqs. \ref{2D} for $R_0=[5.0,15.8,50.0,158.1,500.0] r_G$, an integration time of $10^6 t_G$, a logarithmic temporal resolution of $5 \cdot 10^6$ steps, as in Sect. \ref{sect_1D}, $\lambda_{Edd}=1$ and $v_{z,0}=v_{rot}$. For comparison, we also show the corresponding classic solutions. As expected, the highest differences are observed at smaller radii, where the velocity is higher and the relativistic effects are stronger. We present in Fig. \ref{kepler-2} in the Appendix a detailed plot of the wind dynamics.

Hereafter, we concentrate on the outflow velocity, defined as $v_{out}=\sqrt{v^2_R+v^2_z}$, rather than on the total wind velocity $v=\sqrt{v^2_R+ v^2_{\phi} + v^2_z}$, to better compare our results with observations. The velocity of the observed UFOs is primarily derived through spectroscopy, thanks to the Doppler shift of the wind absorption lines. These lines are usually described with Gaussian or Voight profiles, i.e. they have an average energy and some degree of broadening. The observed wind outflow velocity $v_{obs}$ is usually derived from the average velocity, while the broadening is phenomenologically ascribed to turbulence or rotational motion within the wind.

In our model, $v_{obs}$ is given by the projection of $\vec{v}_R+ \vec{v}_z$ along the line of sight (LOS), while the rotational velocity $v_{\phi}$ only contributes to the broadening of the line, thanks to the axisymmetry of the system. The highest $v_{obs}$ is given by $v_{out}$, and corresponds to the case in which the LOS is parallel to $\vec{v}_R+ \vec{v}_z$. We refer to \citet{Fuku19} for a detailed discussion on this point. Interestingly, the authors discuss the possibility that a rotational motion of the wind around the X-ray corona is responsible for the broadening of the absorption lines, in a similar fashion to the discussion here.

\section{Force multipliers}
\label{sect_frad}
In the accretion disc wind literature, the wind opacity is usually calculated analytically over a broad range of absorption lines in the UV and X-ray energy range (see e.g. \citealp{PK04,RE10,SI10,DP18,Q20}), obtaining the so-called "force multipliers".
Here, we use the radiative transfer code {\it XSTAR} (\citealp{xstar}). \citet{SC11, C09} performed calculations in a similar fashion using Cloudy photoionisation code \citep{Cloudy17}, albeit with a different formalism.

{\it XSTAR} accurately computes the transmitted spectrum $S_T$ through a gas layer, as a function of the following input quantities:
\begin{itemize}
    \item{$S_I$, the incident spectrum, and its integrated ionising luminosity $L_I^{ion}$ in the 1-1000 $Ry$ interval (1 $Ry$= 13.6 $eV$).}
    \item{$r_0$, the distance of the gas from the central luminosity source}
    \item{$n_0$, the gas number density at $r_0$}
    \item{$N_H$, its column density}
    \item{$\alpha$, the coefficient regulating the radial dependence of $n$: $n=n_0\ \Big( \frac{r_0}{r} \Big)^{\alpha}$}
    \item{$v_{broad}$, the gas velocity dispersion regulating the broadening of the absorption features.}
\end{itemize}
The difference $L_I-L_T$ (where $L_I,L_T$ are the integrated luminosity of $S_I,S_T$) corresponds to the amount of radiation absorbed by the wind thanks to its opacity. This difference corresponds to a momentum $\Delta p$ deposited on the wind, which can be written as $\Delta p=\frac{L_I-L_T}{c}=\frac{L_I \cdot f_{rad}}{c}$, where we introduce the new variable $f_{rad}=\frac{L_I-L_T}{L_I}$.
Specifically, for a given set of initial parameters $[S_I,r_0,n_0,N_H,\alpha, v_{broad}]$ we run an {\it XSTAR} simulation and we calculate $f_{rad}$ as:
\begin{equation}
f_{rad}= \frac{\int^{E_1}_{E_0} s_{\nu}^I-s_{\nu}^T }{ \int^{E_1}_{E_0} s_{\nu}^I} d \nu
\end{equation}
where $E_0, E_1$ correspond to the lower and upper energy bound and $s_{\nu}^I, s_{\nu}^T$ are the incident and transmitted flux, respectively, as functions of the frequency $\nu$, so that $\int s_{\nu}^I d \nu = S_I$ and $\int s_{\nu}^T d \nu = S_T$. From a mathematical point of view, $f_{rad}$ corresponds to a weighted average of the wind opacity.
In all the cases of interest, absorption features are comprised in the energy interval between 0.1 $eV$ and $100 keV$, that we choose as $E_0, E_1$, respectively.

It is important to note that {\it XSTAR}, as well as other photoionsation codes such as {\it Cloudy}, does not allow the inclusion of a net velocity of the gas and the related relativistic effects. In order to take them into account, it is convenient to transform the spectra from the rest frame $K$ to the wind reference frame $K'$, and to manipulate $S_I$ and $S_T$, namely to shift their frequencies by a factor $\psi$ and multiply their fluxes by $\psi^3$ (we refer to \citealp{L20} for a detailed explanation). However, since $f_{rad}$ is calculated as a ratio between $S_T$ and $S_I$, the transformations cancel out so that $f_{rad}$ can be directly calculated in $K$ without the need of any relativistic transformation.

We integrate the equation of motion, Eq. \ref{main_eq}, for a radial interval $\Delta r$, in which a wind column density $N_H$ is enclosed, and we include the momentum $\Delta p$. The resulting equation is:
\begin{equation}
\begin{split}
    	\frac{dv}{dt} & =\frac{\Delta p}{4 \pi r^2 m_p N_H} + \frac{L_{bol}'\sigma_T}{4 \pi r^2 c m_p} - \frac{GM}{r^2} \\
	& = \frac{1}{4 \pi r^2 c m_p} \cdot \big(\frac{L_{UX}' \cdot f_{rad}}{N_H}+L_{bol}' \cdot \sigma_T)  - \frac{GM}{r^2}    
\end{split}
\end{equation}
\label{frad_eq}
where we also include Thompson scattering and $L'_{bol},L'_{UX}$ correspond to the bolometric luminosity and the incident luminosity between $E_0$ and $E_1$, respectively. The $'$ symbol indicates that the luminosities are in the $K'$ frame, i. e. $L'_{bol}(L'_{UX})=L_{bol}(L_{UX}) \cdot \frac{1}{\gamma^4 (1+\beta cos(\theta))^4}$. It must be noted that, for this equation to hold, $\Delta r$ must be small enough so that $r$ can be approximated as constant (i.e., $\Delta r \ll r$).

We put ourselves in the same framework of Sect. \ref{sect_2D} (axisymmetry, thin accretion disk and conservation of $l$). The complete set of equations is then:
\begin{subequations}
\begin{align}
	\frac{dv_R}{dt} & =\Big( \frac{\lambda'_{UX} \cdot f_{rad}}{\sigma_T N_H} + \lambda'_{Edd} - 1 \Big) \frac{R}{r^3} + \frac{l^2}{(GM/c)^2 \cdot R^3}\\
	\frac{dv_z}{dt} & =\Big( \frac{\lambda'_{UX} \cdot f_{rad}}{\sigma_T N_H} + \lambda'_{Edd} - 1 \Big) \frac{z}{r^3}\\
	R & = R_0+\int_{t_0}^{t_1} v_R\ dt \\
	z & = \int_{t_0}^{t_1} v_z\ dt \\
    r & = \sqrt{R^2+z^2} \\
    \vec{v}_0 & = (0,v_{rot},v_{z,0}) \\
    \vec{r}_0 & = (R_0,0,0)
\end{align}
\label{2D_M}
\end{subequations}
where, as in Sect. \ref{sect_2D}, $r,t$ are in units of $r_G, t_G$ and $\lambda'_{UX} \equiv L'_{UX}/L_{Edd}, \lambda'_{Edd} \equiv L'_{bol}/L_{Edd}$.

\section{Initial parameters and force multipliers calculation}
\label{n_steps}

Since we are primarily interested in winds from AGN accretion discs, we focus on "typical" values of $\lambda_{Edd}$ between 0.1 and 2.0 and a black hole mass $M=10^8 M_{sun}$ .
However, as we show later in this section, the properties of the wind, and then the values of $f_{rad}$, mainly depend on $\lambda_{Edd}$, rather than on $M$ or $L_{bol}$ alone, so our results are applicable regardless of the black hole mass.

We use the bolometric corrections of \citet{L12} to obtain the 2-10 $keV$ luminosities. We assume a simple powerlaw incident spectrum with photon index $\Gamma=2$, consistent with the typical values observed in AGNs \citep{P05,T11}, to extrapolate $L_{ion}$ and $L_{UX}$.

Regarding the properties of the wind, we concentrate on a set of initial number density $log(n_0/cm^3) \in [10,11,12,13]$ in order to match the ionisation parameters of the observed UFOs, as we will discuss in the following, and a range of $N_H \in [5 \cdot 10^{22}, 10^{24}] cm^{-2}$. As starting radii we use the same set of values of Sect. \ref{sect_1D} and \ref{sect_2D}: $R_0 \in [5.0,15.8,50.0,158.1,500.0] r_G$. In our picture, we assume that all the luminosity comes from a point source located at the coordinate origin ($R=0,\ z=0$). However, we note that observationally the X-ray flux is usually ascribed to a hot "corona", comprised within $\sim 10 r_G$ from the SMBH \citep{Chartas12,Reis13,Reis14,Kara16,CG20,Sz20}, while the UV radiation is due to the disc emissivity which, for a thin disc, has a peak at $\sim 20 r_G$ \citep{Q20}. As a result, we expect that in our code the radiative contribution may not be fully modelled for $R_0<50.0 r_G$; however, we include the cases for $R_0=5.0,15.8 r_G$, since the fate of the wind is governed primarily by $v_0$, rather than by the radiation pressure, as we will show later.

We fix $\alpha$, the exponent regulating the radial dependence of $n$, to 2, so that $n=n_0 \big(\frac{r_0}{r} \big)^2$, as expected by mass conservation for a medium expanding in a spherical geometry. Since we expect a high degree of velocity shear within the wind due to its acceleration, we use a high turbulent velocity $v_{broad}=3000 km\ s^{-1}$ to prevent line saturation. This value is consistent with those typically observed in UFOs (see e.g. \citealp{Fuku19} and references therein).

We briefly summarise here the range of input quantities of our simulations, from which we calculate a grid of $f_{rad}$ values:
\begin{itemize}
    \item{$\lambda_{Edd}$: we divide the interval of interest in four values, logarithmically spaced: $\lambda_{Edd}=[0.1,0.5,1.,2.]$.}
    \item{$n_0$: we divide the range in logarithmic steps: $log(n_0/cm^3) \in [10,11,12,13]$}
    \item{$r_0 \in [5.0,15.8,50.0,158.1,500.0] r_G$.}
    \item{$N_H$: we span the range $[5\cdot 10^{22},10^{24}]\ cm^{-2}$ with steps of $5\cdot 10^{22} cm^{-2}$.}
\end{itemize}

We compute the geometrical thickness of the wind to check whether the $\Delta r \ll r$ condition in Eq. \ref{frad_eq} is met. The results plotted in Fig. \ref{scaling_n} (first three panels) in Appendix show that $\Delta_R /R_0<1$ in all the cases. The only exception is for $R_0=5 r_G, log(n_0/cm^3)=10$ when the column density is very high ($N_H>7\cdot 10^{23} cm^{-2}$). However, as we will discuss later, the properties of the wind are quite independent from $N_H$, and we will focus on the typical UFO value of $N_H=10^{23} cm^{-2}$ \citep{T11} in most of the cases. 
In the last panel of Fig. \ref{scaling_n} we show the ionisation parameter $\xi(r)$, defined as $ \frac{L_{ion}}{n_0 r_0^2}$, for $\lambda_{Edd}=1.0, 0.1$. This parameter is of great importance since the absorption structure of the wind, and then the values of $f_{rad}$, mainly depends on it. Our range of $\xi$, from $\sim 10^0$ to $\sim 10^6$, agrees well with the broad population of UFOs and Warm Absorbers (see e.g. \citealp{Laha14,Sera19}); this, in turn, justifies our range of $log(n_0)$. Moreover, our values for $log(n_0)$ are in agreement with those commonly estimated for the Broad Line Region (BLR) and for UV and X-ray outflows (see e.g. \citealp{elvis00,schurch07,SC11,netzer13}).
Finally, in Fig. \ref{frad_n} we show the values of $f_{rad}$ as a function of the wind parameters.

\subsection{Impact of the initial parameters on the wind dynamics.}
\label{assess}
We solve the system of Eqs. \ref{2D_M} for the whole grid of initial parameters for an integration time of $10^6 t_G$, the same temporal resolution of Sect. \ref{sect_1D} and using a set of $v_{z,0} \in [1.0,10^{-1},10^{-2},10^{-3}]\ v_{rot}$.
For a complete discussion of the results, we first analyse the impact of the different parameters. To characterise the wind dynamics, we define a wind successfully launched if it has a positive (outbound) velocity at the end of the integration time $t=t_1$, and we compute its terminal velocity as $v_t=v_{out} (t_1)=\sqrt{v_R(t_1)^2+v_z(t_1)^2}$.

\begin{figure}
\centering
\includegraphics[width=10cm]{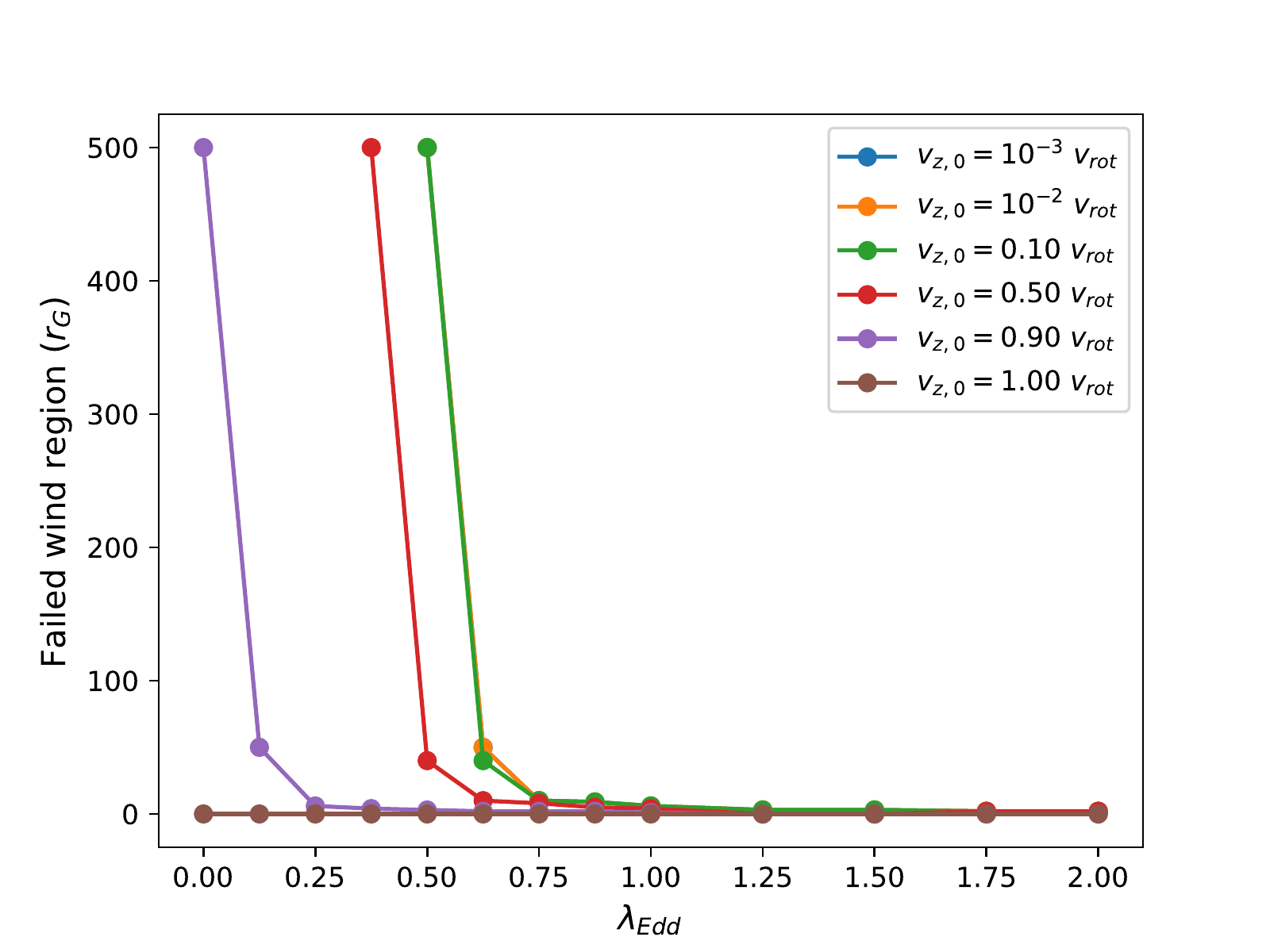}
\caption{Failed wind region as a function of $\lambda_{Edd}=$ for different $v_{z,0}$ (colour coding, see legend). It can be seen that for $v_{z,0} \leq 10^{-1} v_{rot}$ the region becomes quite constant, since the wind dynamics does not depend anymore on the value of $v_{z,0}$. }
\label{corona}
\end{figure}

$\lambda_{Edd}$ and $v_{z,0}$ are the dominant parameters, since they regulate the amount of radiation pressure and the initial velocity of the wind. 
When $v_{z,0}=v_{rot}$, the high initial velocity allows the wind to be successfully launched for any value of $\lambda_{Edd}$ (albeit reaching different $v_t$). 
For $v_{z,0}=10^{-1},10^{-2},10^{-3}\ v_{rot}$, instead, the impact of the initial velocity itself to the wind evolution is almost negligible, and the different solutions are indistinguishable between them. However, $v_{z,0}$ lifts the gas above the disk, displacing it from its equilibrium radius and exposing it to the radiation pressure. The fate of the wind will then depend on $\lambda_{Edd}$ and, secondarily, on $R_0$, but it is unaffected by the exact value of $v_{z,0}$. 
We plot in Fig. \ref{corona} the failed wind region (i.e., the radius up to which the wind cannot be succesfully launched) as a function of $\lambda_{Edd}$, for different $v_{z,0}$. For $v_{z,0}=v_{rot}$, the wind is always successful. Then, the failed region increases for decreasing $v_{z,0}$, and becomes quite constant for $v_{z,0} \leq 10^{-1} v_{rot}$. Please note that in this Figure we do not take into account wind opacity (i.e., we use the same treatment of Sect. \ref{sect_2D}) due to the long computational times required. However, the overall behaviour of $v_{z,0}$ is similar.
For simplicity, from now on we will concentrate only on $v_{z,0}=[10^{-2},1.0] v_{rot}$. We note that $v_{z,0}= v_{rot}$ represents a remarkably high starting velocity, which could be justified only under certain particular physical conditions (see discussion in Sect. \ref{Discussion}) and is significantly higher from the velocities commonly assumed in the literature, which are closer to the $v_{z,0}= 10^{-2} v_{rot}$ case (see e.g. \citealp{PSK00,nomura16,nomura20}). As we will detail in the following, we run the simulations using this starting value as a case study, in order to set an upper bound for the wind velocities that can be reached through radiative pressure, once corrected for relativistic effects.

For what concerns $n_0$, it can give rise to differences in $v_t$ of up to $0.05 c$, but it generally does not affect the overall behaviour of the wind.
Since $f_{rad}$ is always roughly directly proportional to the column density, the wind solutions are also quite independent from $N_H$, because these two terms balance each other in the first term on the right-hand side of Eqs. \ref{2D_M}a,b. We note that this trend holds also for optically-thick winds. In facts, for column densities $\approx \sigma_T^{-1} = 1.7 \cdot 10^{24} cm^{-2}$, line opacity grows equally or less than linearly with $N_H$ \citep{T11}, and so the $f_{rad}/N_H$ term in Eqs. \ref{2D_M}a,b will not contribute more than for the optically-thin wind. We refer to Sect. \ref{n_dependence} in the Appendix for further discussions on $n_0, N_H$. From now on we will focus on $log(n_0/cm^3)=11, N_H=10^{23} cm^{-2}$, if not stated otherwise.

\section{Results}
\label{Results}

\begin{figure*}
\centering
\includegraphics[width=15cm]{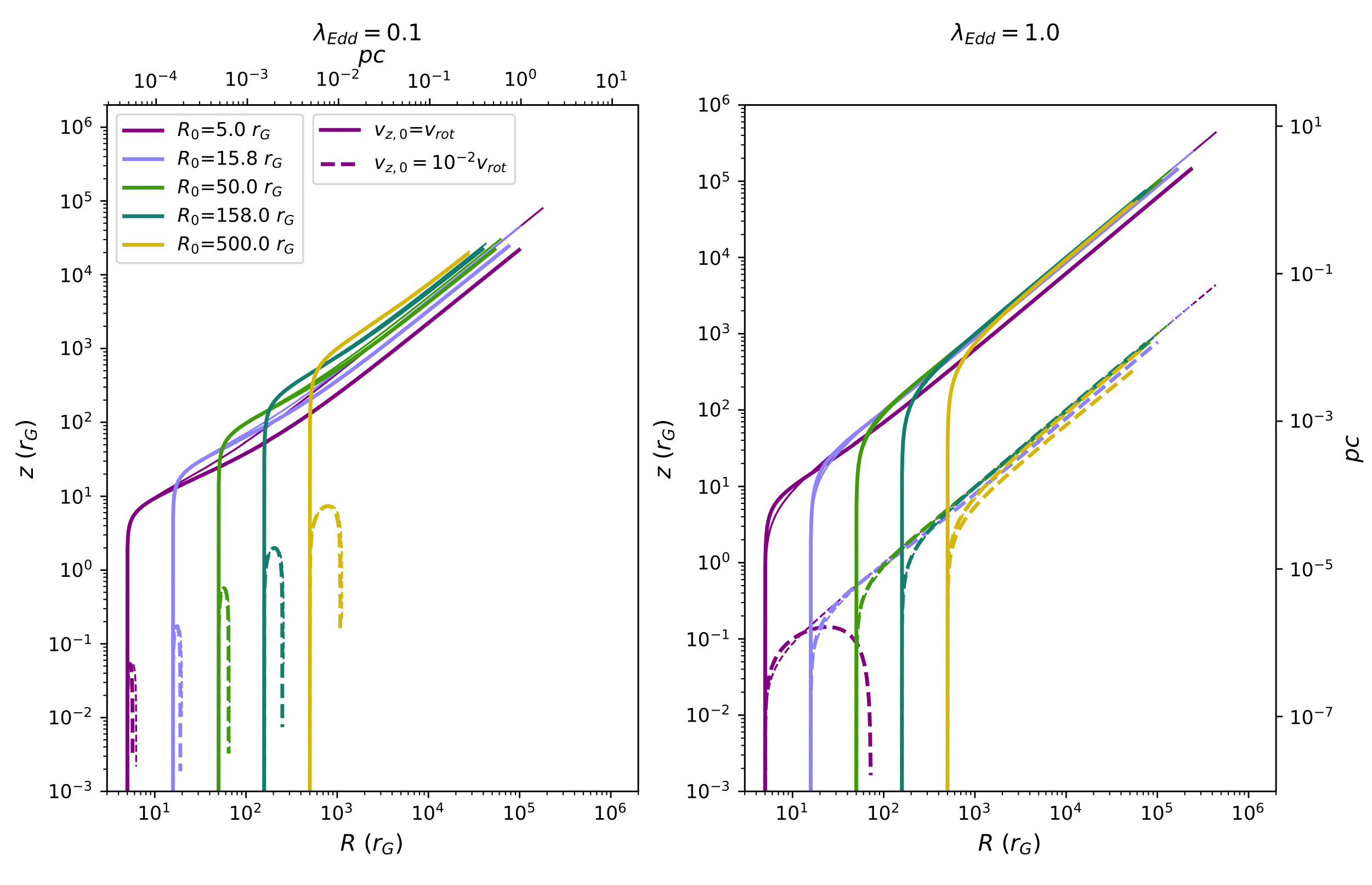}
\caption{Trajectories of the wind in the $z - R$ plane (y and x axis, respectively) for $\lambda_{Edd}=0.1,1.0$ (left and right panel, respectively). Solid(dashed) lines correspond to $v_{z,0}=v_{rot} (10^{-2} v_{rot})$. Thick and thin lines refer to the relativistic and classic (i.e., non-relativistic) treatment. Distances are reported both in units of $r_G$ and $pc$, where the latter are calculated assuming $M=10^8 M_{sun}$.}
\label{r_z}
\end{figure*}

To examine the wind behaviour, we plot in Fig. \ref{r_z} the trajectories in the $R-z$ plane for $\lambda_{Edd}=0.1,1.0$ (left and right panels, respectively). Hereafter, we convert the distance from the black hole (x-axis) from $r_G$ to $pc$ assuming $M=10^8 M_{sun}$. As discussed in Sect. \ref{assess}, the dominant parameter in the wind motion is $v_{z,0}$: for $v_{z,0}= v_{rot}$ (solid lines) the wind is always successful, while when $v_{z,0}= 10^{-2} v_{rot}$ (dashed lines) it can be launched only for $\lambda_{Edd}=1.0$ and $R_0>5 r_G$. In Fig. \ref{r_z} and in the followings, a truncated trajectory corresponds to a failed wind, since we interrupt the numerical integration when the gas falls back to the disk plane. Classic and relativistic trajectories (represented with bold and thin lines, respectively) are almost indistinguishable; however, as we will see in the following, their velocities are rather different.

\begin{figure*}
\centering
\includegraphics[width=15cm]{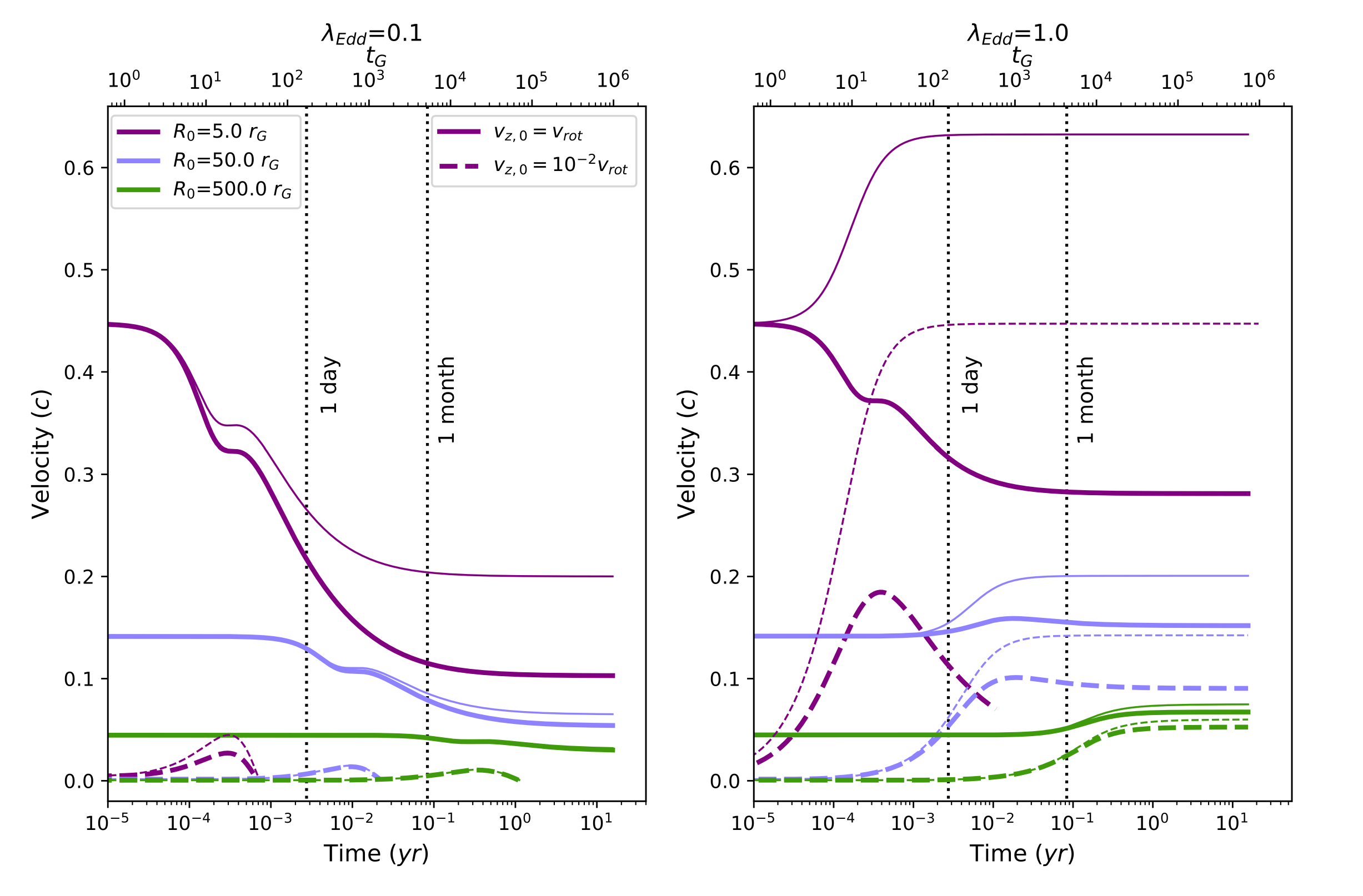}
\caption{Velocity of the wind as a function of $t_G$ for $\lambda_{Edd}=0.1,1.0$ (left and right panel, respectively). Line styles as in Fig. \ref{r_z}. Times are converted from $t_G$ to $yr$ assuming $M=10^8 M_{sun}$.}
\label{v_t}
\end{figure*}

In Fig. \ref{v_t} we show a comparison between the velocities in the relativistic and in the classic treatments (bold and thin lines, respectively) as a function of $t$, for $\lambda_{Edd}=0.1, 1.0$ (left and right panel, respectively). Purple, light blue and green lines corresponds to $R_0= 5.0, 50.0, 500.0 r_G$, respectively. Solid(dashed) lines correspond to $v_{0}=1.0$($=10^{-2} v_{rot}$). 
The impact of relativistic effects is remarkable, especially in the highest velocity cases. The maximum $v_t$ drops from $\approx 0.6$ to less than 0.3 $c$ when taking into account the reduction of radiative pressure due to special relativity effects. We also note that, in most of the cases, the wind attains its terminal velocity within $10^4 t_G$ (corresponding to 0.1 $yr$, i.e. roughly a month, for $M=10^8 M_{sun}$). For $R_0=5 r_G$, this time is reduced to $10^3 t_G$ ($10^{-2} yr$, i.e. a few days). For comparison, we indicate the times corresponding to one day and one month with vertical dotted lines.

\begin{figure*}
\centering
\includegraphics[width=15cm]{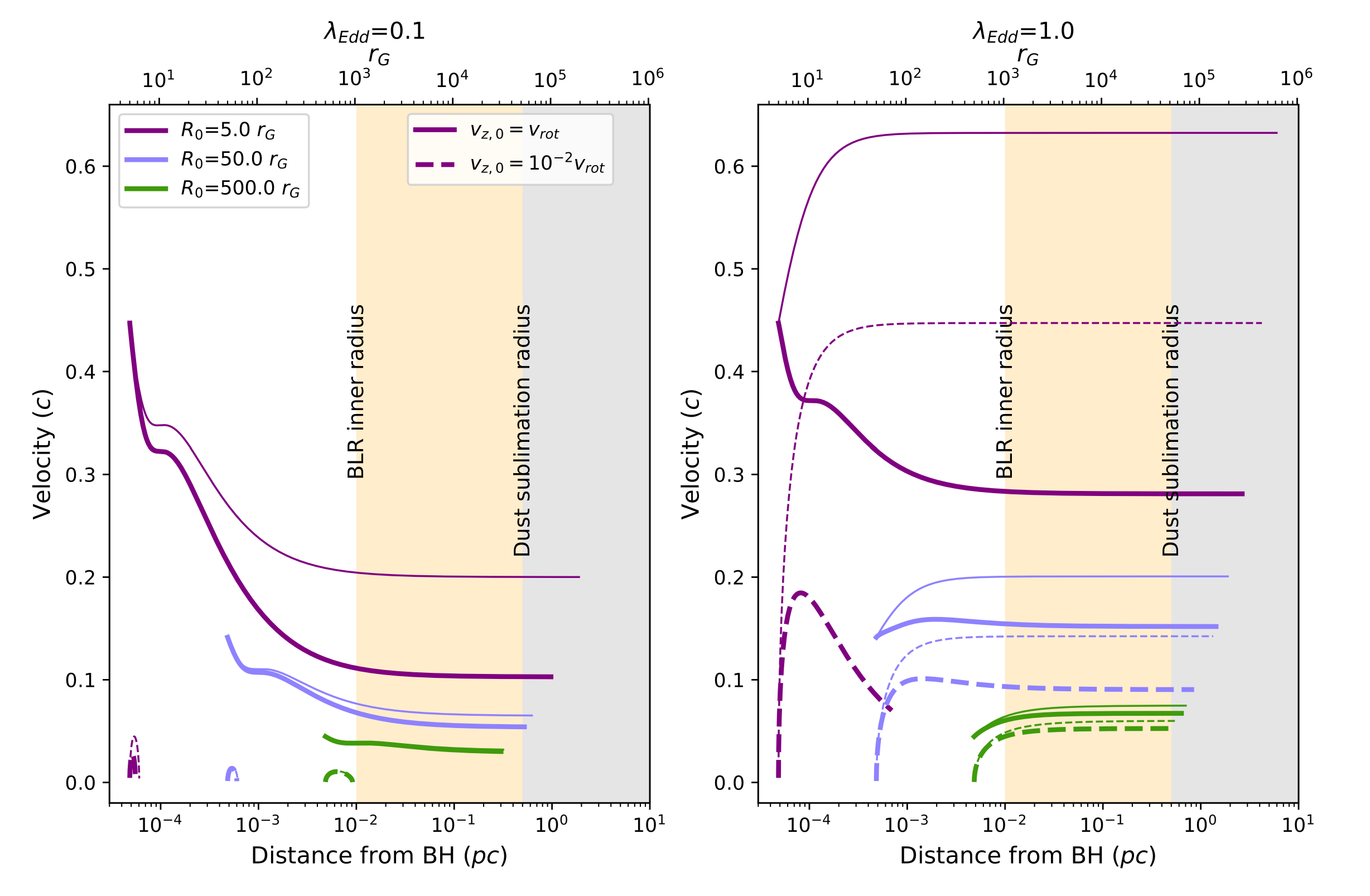}
\caption{Velocity vs. distance from the black hole for $\lambda_{Edd}=0.1,1.0$ (left and right panel, respectively). Line styles as in Fig. \ref{r_z}. Distances are converted from $r_G$ to $pc$ assuming $M=10^8 M_{sun}$.}
\label{v_r}
\end{figure*}

In Fig. \ref{v_r} we show the velocity of the wind as a function of the distance from the black hole. Line styles are as in Fig. \ref{v_t}. To give an idea of the dimensions of the accretion disc - torus system, we also indicate the typical distance at which BLR are observed and the dust sublimation radius, as a proxy of the inner boundary of the torus. We choose 0.01 $pc$ as the BLR inner radius and 0.5 $pc$ as the dust sublimation radius, which marks the boundary between the BLR and the torus (\citealp{CO14,RO14,AD16,ST18,CZ19}). 

For a detailed analysis of the wind dynamics, we show in Fig. \ref{Edd1.0_n} in Appendix the case for $\lambda_{Edd}=1, v_{z,0}=v_{rot}$.

\section{Discussion}
\label{Discussion}

\begin{figure*}
\centering
\includegraphics[width=15cm]{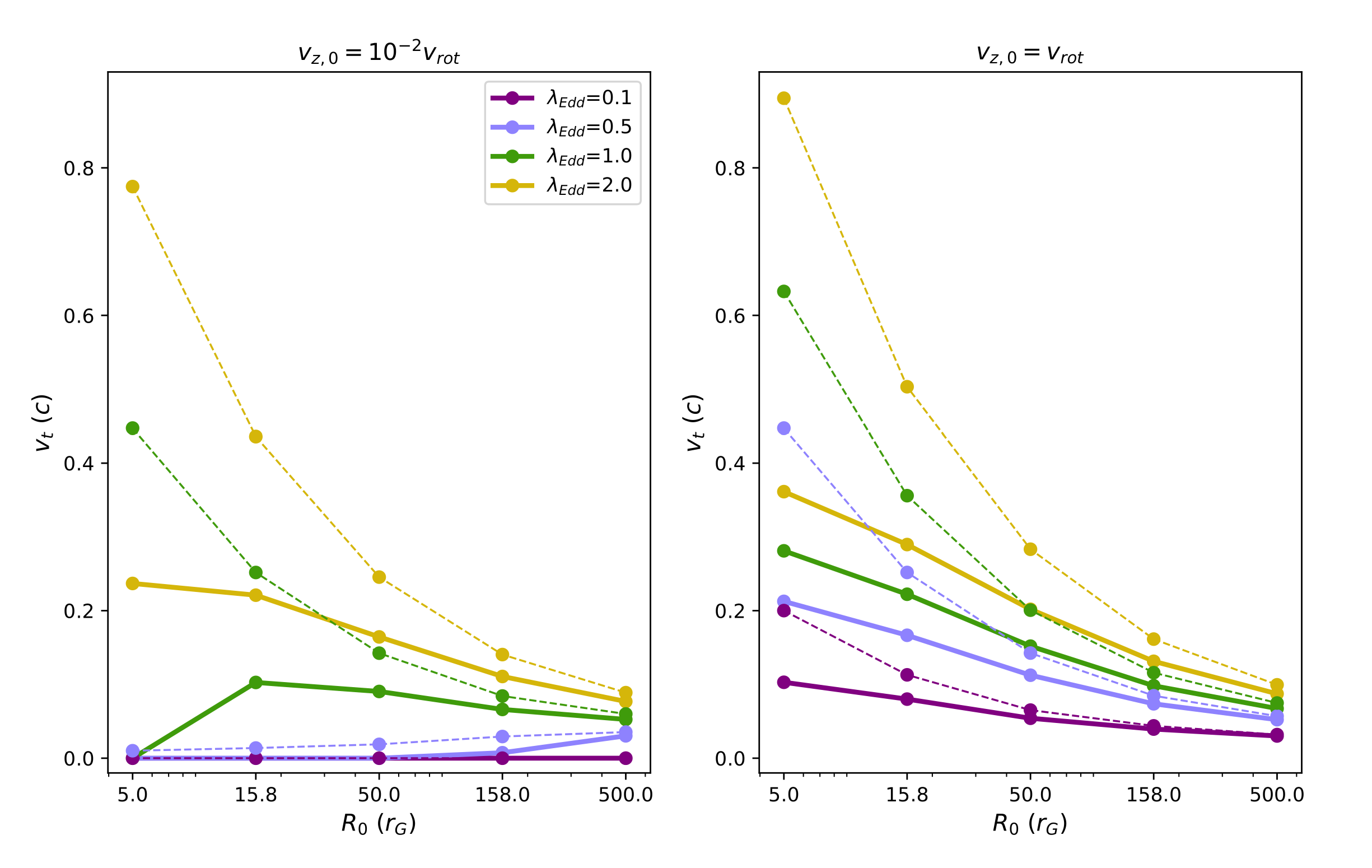}
\caption{Terminal velocities $v_t$ as functions of $R_0$ for different values of $\lambda_{Edd}$ (colour coding). Solid lines refer to the relativistic treatment, dashed to the classic (non-relativistic) one. In the left panel $v_{z,0}= 10^{-2} v_{rot}$, in the right panel $v_{z,0}=v_{rot}$.}
\label{vt_r}
\end{figure*}

To have an idea of the overall behaviour of the wind, we plot (Fig. \ref{vt_r}) $v_t$ as a function of $R_0$ for $v_{z,0}=10^{-2}, 1.0\ v_{rot}$ (left and right panel). We also plot with dashed lines $v_t$ obtained in the classic case. The null values indicate an unsuccessful wind. In order for the wind to attain the typical velocities of UFOs, i.e. $\gtrsim 0.1 c$ (see e.g. \citealp{F17} for a collection of values from the literature), either the luminosity must be very high ($\lambda_{Edd} \geq 1.0$) or the initial velocity $v_{z,0}$, must be comparable to the rotational velocity $v_{rot}$.

\begin{figure*}
\centering
\includegraphics[width=15cm]{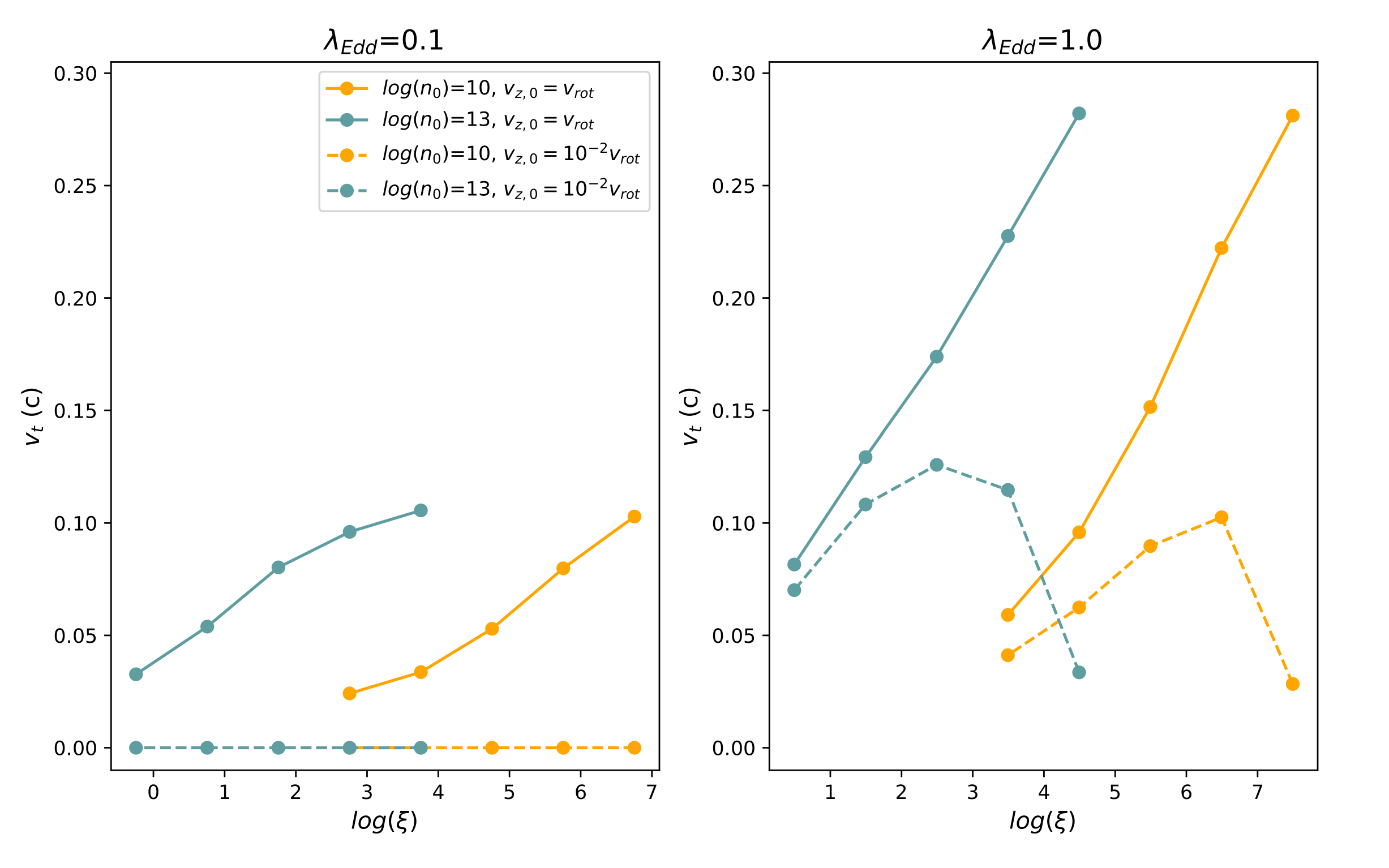}
\caption{$v_t$ as functions of $\xi_0$ for $log(n_0)=10, 13$ (in units of $cm^{-3}$, yellow and gray lines, respectively). Solid and dashed lines refer to $v_{z,0}=v_{rot}$ and $= 10^{-2} v_{rot}$, respectively. In the left panel $\lambda_{Edd}=0.1$, in the right panel =1.0.}
\label{xi_v}
\end{figure*}

In Fig. \ref{xi_v} we plot $v_t$ as a function of $\xi_0$ (which is a monotonically decreasing function of $R_0$) in the densest and in the lightest cases, i.e. $log (n_0/cm^3 \Big)=10$ and $=13$ for $\lambda_{Edd}=0.1$ and =1.0 (left and right panels, respectively). Generally, $v_t$ is found to be a monotonically increasing function of $\xi$ (see e.g. \citealp{T13}). Interestingly, this behaviour can be reproduced only with $v_{z,0} \propto v_{rot}$.

We now compare our results with UFOs from the literature. Our goal is to establish whether the observed UFOs velocities can be reproduced within our radiative driving framework. To do so, we consider two different limiting velocities emerging from our results. The first one is the terminal velocity, $v_t$, which is almost reached by the wind after a very short time, and thus is the most likely to be observed. The second one is the maximum velocity reached by the wind, $v_{max}$, which is associated with the short-living, initial phases of the wind motion (see Figs. \ref{v_t}, \ref{v_r}), and hence represents an upper limit of the observable velocity.
Since we are interested in the highest possible velocities, we focus on the $v_{z,0}=v_{rot}$ cases.

\begin{figure*}
\centering
\includegraphics[width=15cm]{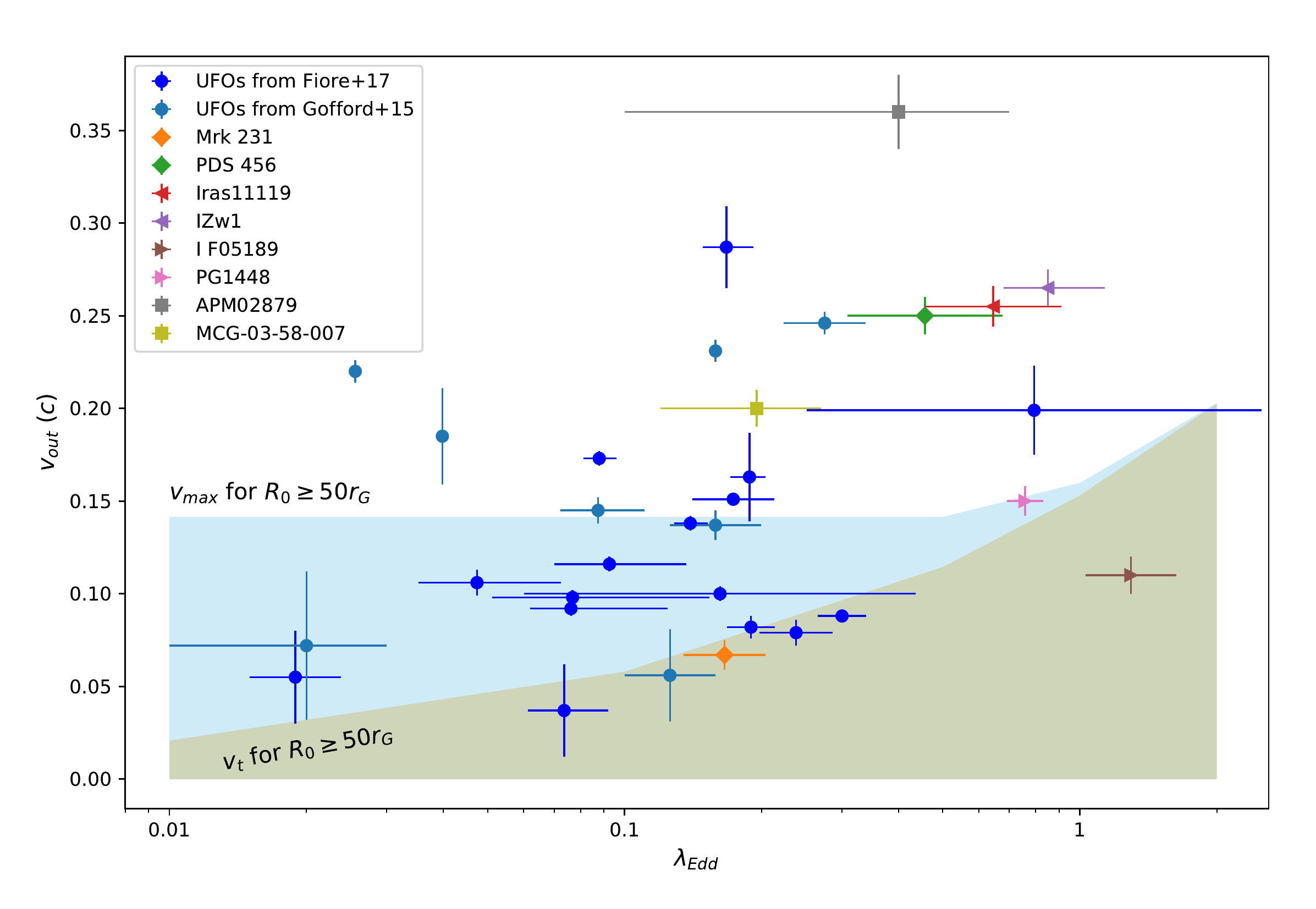}
\caption{Comparison between UFO velocities from the literature (dots and squares) and the results of the present work as a function of $\lambda_{Edd}$. Theoretical limit according to the terminal velocity $v_t$ for $R_0 \geq 50 r_G$ corresponds to the orange shaded area. Limit according to the maximum (but short-lived) velocity $v_{max}$ corresponds to the light blue area. See Sect. \ref{biblio_ref} in Appendix for a description of the sample and the related references.}
\label{UFOs}
\end{figure*}

We compute the highest values of $v_t, v_{max}$ as a function of $\lambda_{Edd}$ for $R_0 \geq 50.0 r_G$, that represents a lower bound for the launching radius of the observed UFOs (see discussion in Sect. \ref{sect_1D}). We plot the results in Fig. \ref{UFOs}, denoting with dark orange and blue the regions below the curves of $v_t, v_{max}$, respectively. These regions correspond to the allowed velocity ranges.

We show with different symbols (see legend) the location of several UFOs reported in the literature
(see Sect. \ref{biblio_ref} in Appendix for references on the single sources). Interestingly, many points lie above the $v_t$ (orange) allowed region, and many also above the $v_{max}$ (blue) region.  Such high velocities, which can be hardly explained within our radiative driving model, may lend support to other launching mechanisms. As discussed in Sect. \ref{assess}, $v_{z,0}=v_{rot}$ represents an upper bound of the expected launching velocities for a radiatively-driven wind. Lower, physically-motivated $v_{z,0}$ would result in even lower limiting velocities than those reported in Fig. \ref{UFOs}, thus strengthening our conclusion that radiative driving, once corrected for special relativity effects, is not sufficient to produce the observed UFO velocity distribution.

Particularly, we signal the possibility of a magnetocentrifugal acceleration mechanism, which is capable to drive the wind up to very high terminal velocities. Typical values are $\sim 1-3$ times the rotational velocity at the wind launching radius $R_0$ (\citealp{F10,F14,T13,Cui20}). For $R_0=50 r_G$, this corresponds to terminal velocities between 0.14 and 0.42 $c$, thus easily accounting for the observed UFO velocities.

We outline two interesting implications of our results. 
Several observations show the simultaneous presence in AGN X-ray spectra of fast absorbers with comparable $v_{out} \sim 0.1- 0.2 c$ and orders of magnitude differences between their $\xi$ \citep{Long15, Sera19, Reeves20}. This evidence can be easily explained within our model. 
The weak dependence of the outflow solutions from $n_0$ indicates that different wind elements can be launched with similar velocity (and column density) but rather different ionisation parameters, as shown in Fig. \ref{xi_v}. This, in turn, is due to the sub-dominant contribution of the force multipliers (and then of the line driving) with respect to Thompson scattering, as outlined also in \citet{Dannen19}.

Secondly, failed winds (FW) are a natural outcome in any radiative driven scenario, and we expect their presence to be ubiquitous, if the radiation is the main driver. In fact, in our analysis we show that successful winds can be launched only through very high launching velocities ($v_{z,0} \propto v_{rot}$) or extreme luminosities $\lambda_{Edd}$ ($\gtrsim 1$). However, many AGNs hosting UFOs have $\lambda_{Edd} \sim 0.1$ (see Fig. \ref{UFOs}), and such high $v_{z,0}$ are very difficult to justify in the framework of a steady-state accretion disc, unless postulating a "kick velocity" through hydrodynamic instabilities (see e.g. \citealp{Jan11} and references therein), disc magnetic reconnection \citep{Dimatteo98,E20,Ripp20} or, again, resorting to a large scale MHD-driven outflow (see \citealp{Yuan15} for numerical simulations). Within the dynamics of the accretion-ejection system, we expect the FW to act as a shield for the gas launched at higher $R_0$, possibly regulating its ionisation status and observational properties \citep{Gius19,Gius20}. However, we note that the effectiveness of FW in favouring the launching of more distant layers of gas has not been proven yet (\citealp{H14}; see discussion in \citealp{Zapp20} and references therein). FW are confined into a narrow equatorial region (i.e., their $z$ height is $\ll$ than the radial coordinate $r$, see Fig. \ref{r_z}), since the gravitational attraction prevents them from reaching high altitudes before bouncing back to the accretion disc, making them particularly difficult to observe. A careful modelisation of the wind duty cycle and of the disc region is needed in order to further shed light on this topic.

Finally, we note that our results are robust also in case of X-ray luminosity variability, as observed in several sources (see e.g. \citealp{N15,P17}). As discussed above, for the typical UFO ionisation degrees most of the radiative pressure is channelled through Thompson scattering, rather than line pressure. Thus, the driving luminosity is $L_{bol}$ (expressed here as $\lambda_{Edd}$), that significantly varies in AGNs only on timescales larger than tens of years, rather than the X-ray luminosity (expressed through $\lambda_{UX}$).

\section{Conclusion}
\label{conclusion}
Special relativity effects strongly reduce the radiative pressure exerted on fast moving clumps of gas, as in the case of UFOs from accretion discs, as well as BAL winds, as discussed in the Introduction. In our work, we carried out an extensive analysis of the radiative driving for a disc wind accounting for these effects. Our main findings can be summarised as follows:
\begin{itemize}
\item{The dynamics of the wind is primarily governed by the AGN luminosity and the launching velocity $v_{z,0}$. For high luminosity, $\lambda_{Edd}=1.0$, the wind is successfully launched independently from $v_{z,0}$, while for $\lambda_{Edd}=0.1$ a higher $v_{z,0}$, of the order of the disc rotational velocity, is required in order to overcome the gravitational attraction from the central black hole (see Fig. \ref{r_z}).}
\item{Shortly after the launch of the wind (between one day and one month, depending on $R_0$), the wind attains a roughly constant velocity $v_t$, which is conserved until the end of the integration time ($10^6 t_G$, i.e. $10 yr$ for a black hole mass of $10^8 M_{sun}$). After $\sim$ 1 month, the wind reaches BLR-like distances, possibly suggesting an interaction with the gas in the BLR orbiting above the accretion disc.}
\item{The inclusion of special relativity effects reduces the radiative pressure exerted on the wind. This, in turn, leads to remarkably lower $v_t$ with respect to the classical treatment, up to 50 \% less for winds launched at the smallest $R_0$. Within the relativistic treatment, we find an upper limit of $v_t=0.15 c$ for the highest luminosity case ($\lambda_{Edd}=1.0$) and a launching radius $ \geq 50 r_G$, in agreement with the observed UFO locations.}
\item{Interestingly, we find that most of the UFO velocities from the literature cannot be reproduced within our radiative driving scenario. For the majority of the sources, which have $\lambda_{Edd}$ between 0.03 and 1, the luminosity is too low to reproduce the observed $v_{out}$. This evidence suggests that other acceleration mechanisms are at play. In particular, we suggest the possibility of magnetic driving, which could easily account for the observed $v_{out}$.}
\end{itemize}

\emph{Acknowledgements.} We thank the referee for the valuable comments that helped improve the paper and Manuela Bischetti for discussions on the UFO host galaxies. AL deeply thanks all the staff at the Harvard \& Smithsonian Center for Astrophysics for their warm welcome during the time spent there. AL, EP, FT, LZ  acknowledge financial support under ASI-INAF contract 2017-14-H.0.
EP, FF acknowledge support from PRIN MIUR project "Black Hole winds and
the Baryon Life Cycle of Galaxies: the stone-guest at the galaxy evolution
supper", contract \#2017PH3WAT

\begin{appendix}
\label{appendix}

\section{Three-dimensional wind}
\begin{figure*}
\centering
\includegraphics[width=15cm]{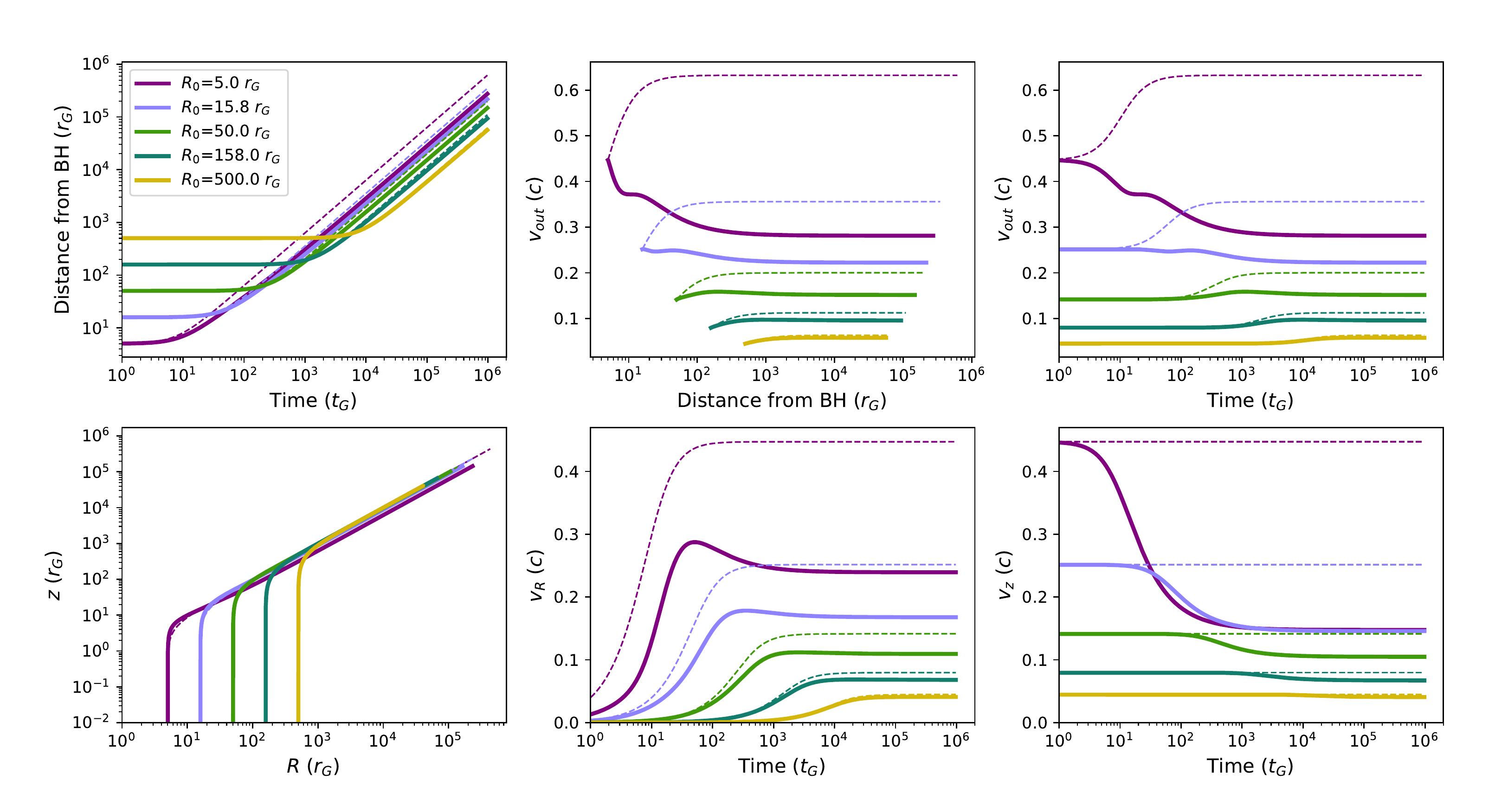}
\caption{Wind trajectories for $v_{z,0}=v_{rot}$ and $\lambda_{Edd}=1$. From top to bottom and left to right: distance $r$ from the BH (where $r = \sqrt{R^2+z^2}$) as a function of $t$; outflow velocity $v_{out}$ as a function of $r$ ($v_{out} = \sqrt{v_R^2+v_z^2}$); $v_{out}$ as a function of $t$; trajectories of the wind in the $z-R$ plane; $v_R$ (radial velocity) as a function of $t$; $v_z$ (vertical velocity) as a function of $t$. Solid(dashed) lines refer to the relativistic(classic) treatment.}
\label{kepler-2}
\end{figure*}

We show in Fig. \ref{kepler-2} a detailed analysis of the wind trajectories for the preliminary three-dimensional model presented in Sect. \ref{sect_2D}.

\section{Geometrical and physical properties of the wind}
\begin{figure*}
\centering
\includegraphics[width=15cm]{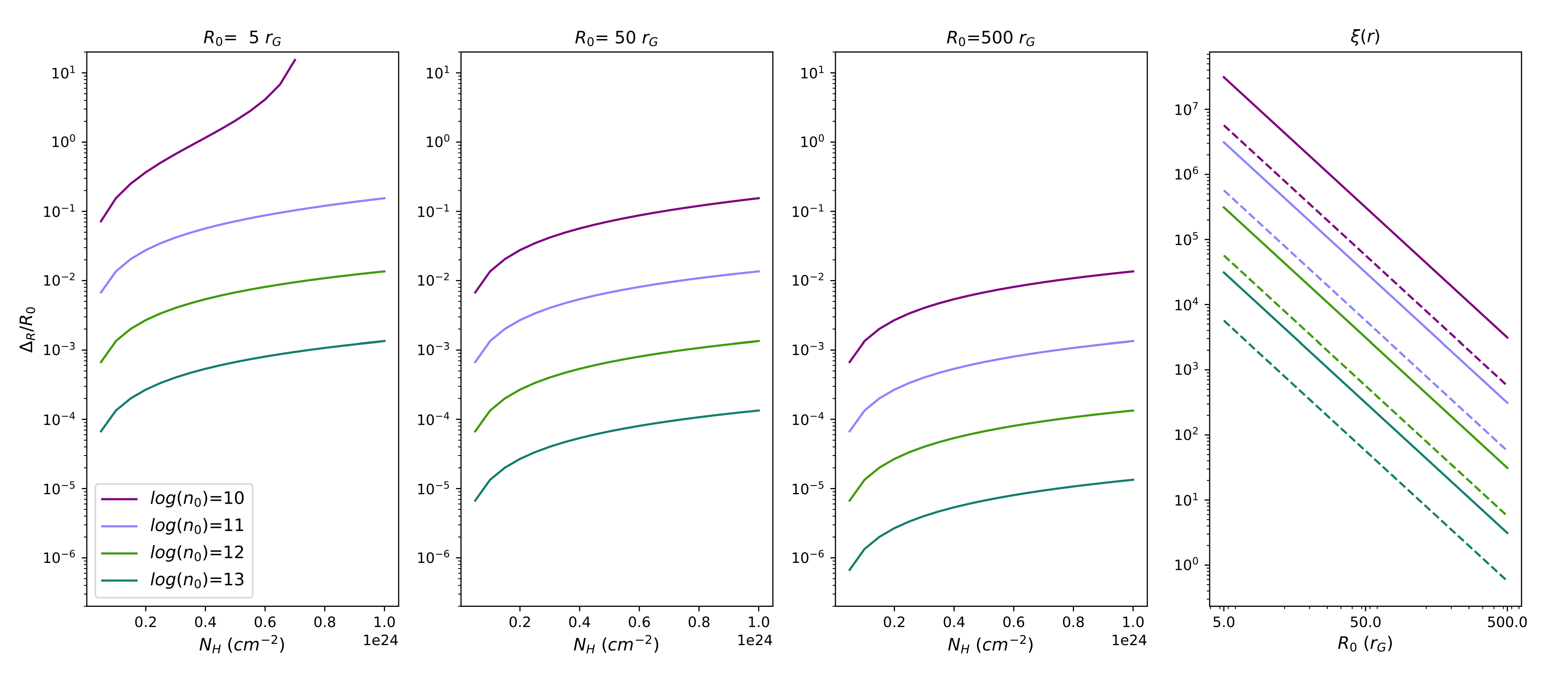}
\caption{First three panels: geometrical thickness of the wind as a function of $N_H$ for $log(n_0)=10,11,12, 13$ (in units of $cm^{-3}$, colour coding) and increasing $R_0$ (from left to right). Last panel: ionisation parameter as a function of $R_0$ for different $log(n_0)$ (colour coding) and $\lambda_{Edd}=0.1, 1.0$ (dashed and solid lines, respectively).}
\label{scaling_n}
\end{figure*}

We show in Fig. \ref{scaling_n} (first three panels) the geometrical thickness of the wind.
With our density profile $n(r)=n_0 \big( \frac{r_0}{r} \big)^2$ (see Sect. \ref{n_steps}), the upper limit for the column density corresponds to $N_{H,max}=n_0 r_0$.
For all the values of $n_0, r_0$ in this paper, $N_{H,max}>10^{24} cm^{-2}$. The only exception is for $n_0=10^{10} cm^{-3},r_0=5 r_G$ (first panel, blue line), for which $N_{H,max}=7.4 \cdot 10^{23} cm^{-2}$. 
In the last panel of Fig. \ref{scaling_n} we plot $\xi(r)$ for $\lambda_{Edd}=0.1, 1.0$.

\begin{figure*}
\centering
\includegraphics[width=18cm]{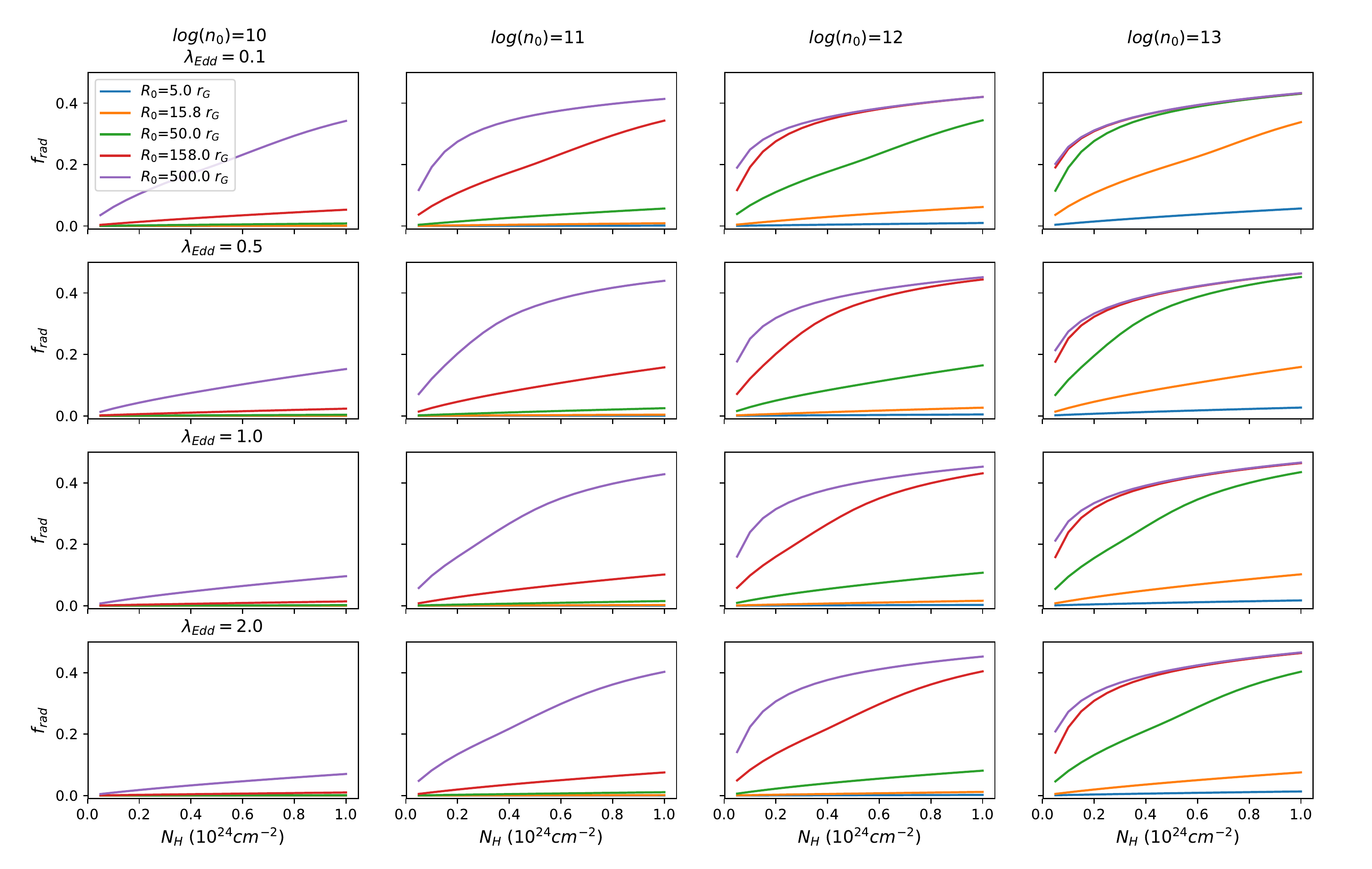}
\caption{Values of \frad\ as functions of $N_H$ for $log(n_0)=10,11,12, 13$ (in units of $cm^{-3}$, from left to right) and $\lambda_{Edd}=0.1,0.5,1.0,2.0$ (from top to bottom). }
\label{frad_n}
\end{figure*}

Fig. \ref{frad_n} shows \frad\ as a function of $N_H$, for different $R_0$ (colour coding) and $log(n_0/cm^3)=10,11,12,13$ (from left to right). From top to bottom, $\lambda_{Edd}=0.1,0.5,1.0,2.0$.

\section{Dependence of the wind dynamics on $n_0, N_H$}
\label{n_dependence}
\begin{figure*}
\centering
\includegraphics[width=15cm]{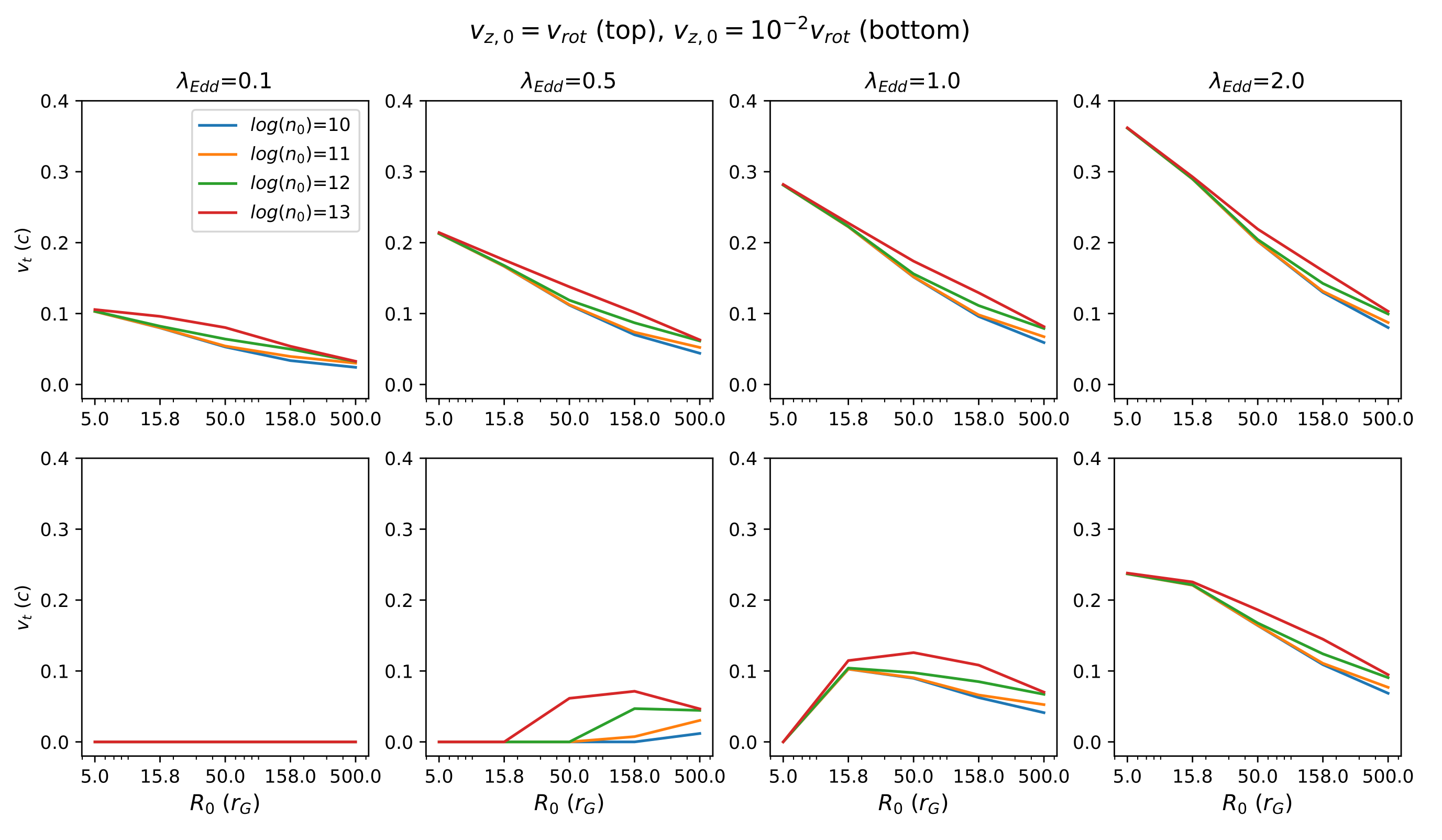}
\caption{Terminal velocity of the wind for $log(n_0)=10,11,12, 13$ (in units of $cm^{-3}$, colour coding) and for different luminosities (from left to right $\lambda_{Edd}=0.1,0.5,1.0,2.0$). In the top(bottom) panel $v_{z,0}=v_{rot}$($v_{z,0}=10^{-2} v_{rot}$). For simplicity, we fix $N_H=10^{23} cm^{-2}$. }
\label{n_impact}
\end{figure*}

To give an idea of the impact of $n_0$ on the wind dynamics, we show in Fig. \ref{n_impact} the terminal velocity $v_t$ as a function of $R_0$ for different $n_0$ (colour coding). From left to right, $\lambda_{Edd}=0.1,0.5,1.0,2.0$. Top (bottom) row refers to $v_{z,0}=v_{rot}$($v_{z,0}=10^{-2} v_{rot}$). $v_t$ is similar for any value of $n_0$, except for $v_{z,0}=0.01 v_{rot}$ and $\lambda_{Edd}=0.5,1.0$, where the initial velocity is low enough (and the luminosity is not too low nor too high) so that different $f_{rad}$ values can give rise to different wind behaviours.

\begin{figure*}
\centering
\includegraphics[width=15cm]{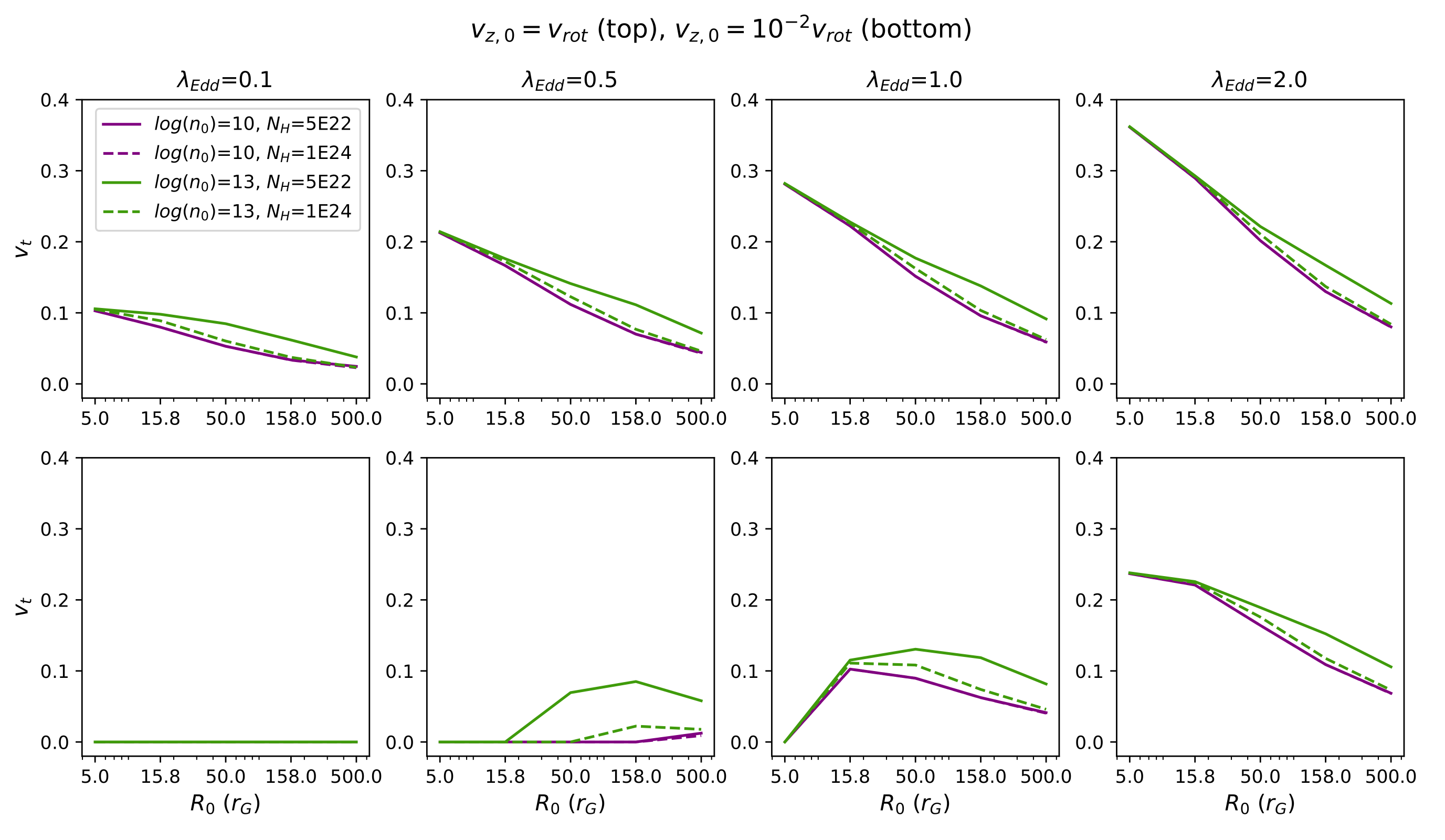}
\caption{Terminal velocity of the wind for $log(n_0)=10,13$ (in units of $cm^{-3}$, purple and green lines) and $N_H=5\cdot 10^{22}$ and $=10^{24}$ (in units of $cm^{-2}$, solid and dashed lines). In the top(bottom) panel $v_{z,0}=v_{rot}$($v_{z,0}=10^{-2} v_{rot}$).}
\label{dual_nh}
\end{figure*}

Similarly, in Fig. \ref{dual_nh} we plot $v_t$ for $log(n_0/cm^3)=10,13$ for the lower and upper values of $N_H$ ($5\cdot 10^{22}$ and $10^{24}cm^{-2}$). It can be seen that the overall dynamics of the wind is weakly sensitive to $N_H$.

\section{Results}
\begin{figure*}
\centering
\includegraphics[width=15cm]{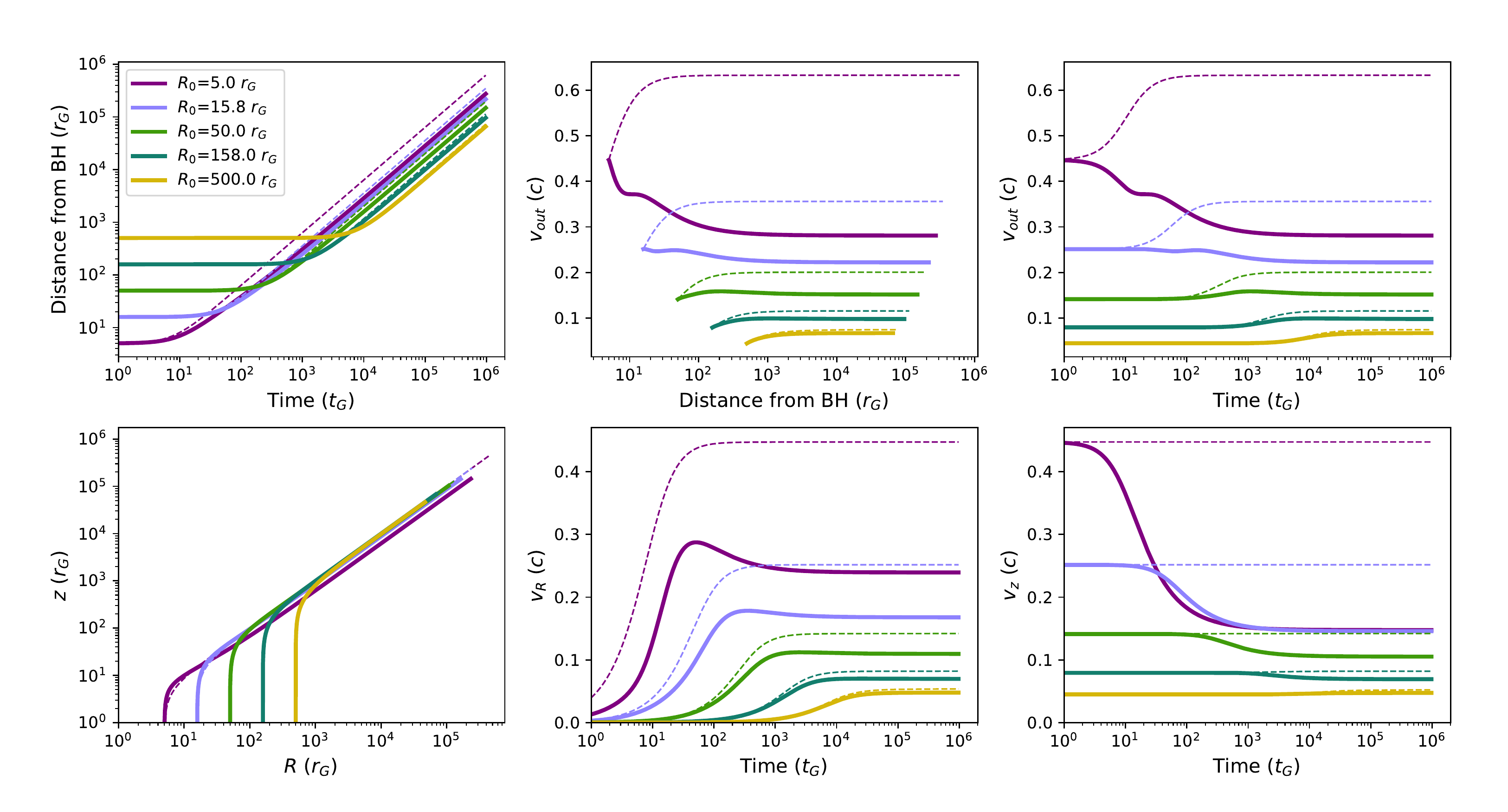}
\caption{Wind solutions in the relativistic and classic cases (solid and dashed lines, respectively) for $\lambda_{Edd}=1, v_{z,0}=v_{rot}$ and taking into account the wind opacity. Meaning of the panels as in Fig. \ref{kepler-2}}
\label{Edd1.0_n}
\end{figure*}

We present in Fig. \ref{Edd1.0_n} a detailed analysis of the wind dynamics for $\lambda_{Edd}=1, v_{z,0}=v_{rot}, log(n_0/cm^3)=11, N_H=10^{23} cm^{-2}$ and taking into account the wind opacity (Sect. \ref{sect_frad}).

\section{Notes on UFOs sources from literature}
\label{biblio_ref}
We show in Table \ref{UFOsample} the properties of the sources plotted in Fig. \ref{UFOs}.
The sample is composed of three main groups. The first one is taken from the \cite{T11} sample, with the updated $L_{bol}$ from \cite{F17}, while the second one is from \cite{Goff15}. The third one comprises the sources reported in \cite{F17} and not previously reported in \cite{T11}, together with other individually-reported UFOs published from 2017 on for which robust estimates of both $\lambda_{Edd}$ and $v_{out}$ are available. Where possible, the black hole mass $M$ and the AGN bolometric luminosity $L_{bol}$ have been updated with recent works, listed in the last column.

\begin{table*}[t]
\centering
\begin{tabular}{c l c c c c}
\hline
Group & Name & $\lambda_{Edd}$ & $v$ ($c$) & Refs. \\
\hline\hline
 & NGC4151 & $0.05\pm 0.01$ & $0.106\pm0.007$ & A \\
 & IC4329A & $0.08\pm 0.03$ & $0.098\pm0.004$ \\
 & Mrk509 & $0.09\pm 0.01$ & $0.173\pm0.004$ & A \\
 & Mrk509 & $0.14\pm 0.01$ & $0.138\pm0.004$ & A \\
 & ARK120 & $0.17\pm 0.02$ & $0.29\pm0.02$ & A \\
 & Mrk79 & $0.08\pm 0.01$ & $0.092\pm0.004$ & B \\
 & NGC4051 & $0.07\pm 0.01$ & $0.04\pm0.02$ & B \\
 & Mrk766 & $0.19\pm 0.02$ & $0.082\pm0.006$ & C \\
1 & Mrk766 & $0.30\pm 0.03$ & $0.088\pm0.002$ & C \\
 & Mrk841 & $0.02\pm 0.00$ & $0.06\pm0.02$ & D \\
 & 1H0419-577 & $0.24\pm 0.04$ & $0.079\pm0.007$ & E \\
 & Mrk290 & $0.19\pm 0.02$ & $0.16\pm0.02$ & B \\
 & Mrk205 & $0.16\pm 0.10$ & $0.100\pm0.004$ & F \\
 & PG1211+143 & $0.17\pm 0.03$ & $0.151\pm0.003$ & D \\
 & MCG-5-23-16 & $0.09\pm 0.02$ & $0.116\pm0.004$ & G \\
 & NGC4507 & $0.79\pm 0.54$ & $0.20\pm0.02$ \\
\hline
 & 3C111 & $0.02\pm 0.01$ & $0.07\pm0.04$ & H \\
 & 3C3903 & $0.09\pm 0.02$ & $0.145\pm0.007$ & A \\
 & 4C+74.26 & $0.04\pm 0.00$ & $0.18\pm0.03$ \\
2 & ESO103-G035 & $0.13\pm 0.03$ & $0.06\pm0.03$ \\
 & MR2251-178 & $0.16\pm 0.03$ & $0.137\pm0.008$ \\
 & Mrk279 & $0.03\pm 0.00$ & $0.220\pm0.006$ & B \\
 & NGC5506 & $0.28\pm 0.05$ & $0.246\pm0.006$ & I \\
 & SQJ2127 & $0.16\pm 0.00$ & $0.231\pm0.006$ \\
\hline
 & Mrk231 & $0.17\pm 0.03$ & $0.067\pm0.008$ & L,M \\
 & PDS456 & $0.46\pm 0.15$ & $0.25\pm0.01$ & N \\
 & Iras 11119 & $0.65\pm 0.19$ & $0.26\pm0.01$ & L,O \\
3 & IZwicky1 & $0.85\pm 0.17$ & $0.27\pm0.01$ & P \\
 & Iras 05189 & $1.30\pm 0.27$ & $0.110\pm0.010$ & Q,R \\
 & PG1448 & $0.76\pm 0.077$ & $0.150\pm0.008$ & S \\
 & APM08279 & $0.40\pm 0.30$ & $0.36\pm0.02$ & T,U \\
 & MCG-03-58-007 & $0.20\pm 0.07$ & $0.20\pm0.01$ & V \\
\hline
\end{tabular}
\caption{Sources plotted in Fig. \ref{UFOs}. From left to right, columns indicate the number of the group, the source name, $\lambda_{Edd}$, the UFO velocity $v_{out}$ (in units of $c$) and the additional references for each source. Group 1: UFO $v_{out}$ and black hole mass $M$ from \cite{T12}, $L_{bol}$ from \cite{F17}. Group 2: values from \cite{Goff15}. For both groups, $M, L_{bol}$ have been updated (where possible) with recent values from the literature, see reference column. Group 3: individual sources, see reference column. References: A \cite{peterson04}, B \cite{ricci17}, C \cite{bentz18}, D \cite{vestergaard06}, E \cite{tilton13}, F \cite{kelly07}, G \cite{caglar20}, H \cite{chatt11}, I \cite{ricci17m}, L \cite{N18}, M \cite{Feruglio15}, N \cite{N15}, O \cite{T15}, P \cite{Reeves19}, Q \cite{Smith19}, R \cite{onori17}, S \cite{La20}, T \cite{C09}, U \cite{Saturni18}, V \cite{Braito18} }
\label{UFOsample}
\end{table*}

\end{appendix}

\begin{thebibliography}{}
\bibitem[Adhikari et al.(2016)]{AD16} Adhikari, T.~P., R{\'o}{\.z}a{\'n}ska, A., Czerny, B., et al.\ 2016, \apj, 831, 68
\bibitem[Bentz \& Manne-Nicholas(2018)]{bentz18} Bentz, M.~C. \& Manne-Nicholas, E.\ 2018, \apj, 864, 146. doi:10.3847/1538-4357/aad808
\bibitem[Bischetti et al.(2019)]{Bischetti19} Bischetti, M., Piconcelli, E., Feruglio, C., et al.\ 2019, \aap, 628, A118
\bibitem[Blandford \& Payne(1982)]{BP82} Blandford, R.~D., \& Payne, D.~G.\ 1982, \mnras, 199, 883
\bibitem[Braito et al.(2018)]{Braito18} Braito, V., Reeves, J.~N., Matzeu, G.~A., et al.\ 2018, \mnras, 479, 3592
\bibitem[Bruni et al.(2019)]{bruni19} Bruni, G., Piconcelli, E., Misawa, T., et al.\ 2019, \aap, 630, A111
\bibitem[Caballero-Garcia et al.(2020)]{CG20} Caballero-Garcia, M.~D., Papadakis, I.~E., Dovciak, M., et al.\ 2020, arXiv e-prints, arXiv:2007.00597
\bibitem[Caglar et al.(2020)]{caglar20} Caglar, T., Burtscher, L., Brandl, B., et al.\ 2020, \aap, 634, A114. doi:10.1051/0004-6361/201936321
\bibitem[Chartas et al.(2012)]{Chartas12} Chartas, G., Kochanek, C.~S., Dai, X., et al.\ 2012, \apj, 757, 137
\bibitem[Chartas et al.(2009)]{C09} Chartas, G., Saez, C., Brandt, W.~N., et al.\ 2009, \apj, 706, 644
\bibitem[Chatterjee et al.(2011)]{chatt11} Chatterjee, R., Marscher, A.~P., Jorstad, S.~G., et al.\ 2011, \apj, 734, 43. doi:10.1088/0004-637X/734/1/43
\bibitem[Coffey et al.(2014)]{CO14} Coffey, D., Longinotti, A.~L., Rodr{\'\i}guez-Ardila, A., et al.\ 2014, \mnras, 443, 1788\bibitem[Czerny(2019)]{CZ19} Czerny, B.\ 2019, Open Astronomy, 28, 200
\bibitem[Contopoulos \& Lovelace(1994)]{CL94} Contopoulos, J., \& Lovelace, R.~V.~E.\ 1994, \apj, 429, 139
\bibitem[Cui \& Yuan(2020)]{Cui20} Cui, C., \& Yuan, F.\ 2020, \apj, 890, 81
\bibitem[Dannen et al.(2019)]{Dannen19} Dannen, R.~C., Proga, D., Kallman, T.~R., et al.\ 2019, \apj, 882, 99
\bibitem[Di Matteo(1998)]{Dimatteo98} Di Matteo, T.\ 1998, \mnras, 299, L15
\bibitem[Di Matteo et al.(2005)]{Dimatteo05} Di Matteo, T., Springel, V., \& Hernquist, L.\ 2005, \nat, 433, 604
\bibitem[Dyda \& Proga(2018)]{DP18} Dyda, S., \& Proga, D.\ 2018, \mnras, 481, 5263
\bibitem[Elvis(2000)]{elvis00} Elvis, M.\ 2000, \apj, 545, 63
\bibitem[Ergun et al.(2020)]{E20} Ergun, R.~E., Ahmadi, N., Kromyda, L., et al.\ 2020, \apj, 898, 154
\bibitem[Ferland et al.(2017)]{Cloudy17} Ferland, G.~J., Chatzikos, M., Guzm{\'a}n, F., et al.\ 2017, \rmxaa, 53, 385
\bibitem[Feruglio et al.(2015)]{Feruglio15} Feruglio, C., Fiore, F., Carniani, S., et al.\ 2015, \aap, 583, A99
\bibitem[Fiore et al.(2017)]{F17} Fiore, F., Feruglio, C., Shankar, F., et al.\ 2017, \aap, 601, A143
\bibitem[Fukumura et al.(2010)]{F10} Fukumura, K., Kazanas, D., Contopoulos, I., et al.\ 2010, \apj, 715, 636
\bibitem[Fukumura et al.(2014)]{F14} Fukumura, K., Tombesi, F., Kazanas, D., et al.\ 2014, \apj, 780, 120
\bibitem[Fukumura \& Tombesi(2019)]{Fuku19} Fukumura, K., \& Tombesi, F.\ 2019, \apjl, 885, L38
\bibitem[Gaspari et al.(2011)]{Ga11} Gaspari, M., Melioli, C., Brighenti, F., et al.\ 2011, \mnras, 411, 349
\bibitem[Gofford et al.(2015)]{Goff15} Gofford, J., Reeves, J.~N., McLaughlin, D.~E., et al.\ 2015, \mnras, 451, 4169
\bibitem[Giustini \& Proga(2019)]{Gius19} Giustini, M. \& Proga, D.\ 2019, \aap, 630, A94
\bibitem[Giustini \& Proga(2020)]{Gius20} Giustini, M. \& Proga, D.\ 2020, arXiv:2002.07564
\bibitem[Hagino et al.(2015)]{Hagino15} Hagino, K., Odaka, H., Done, C., et al.\ 2015, MNRAS, 446, 663
\bibitem[Hamann et al.(2018)]{hamann18} Hamann, F., Chartas, G., Reeves, J., et al.\ 2018, \mnras, 476, 943
\bibitem[Higginbottom et al.(2014)]{H14} Higginbottom, N., Proga, D., Knigge, C., et al.\ 2014, ApJ, 789, 19
\bibitem[Hopkins \& Elvis(2010)]{HE10} Hopkins, P.~F. \& Elvis, M.\ 2010, \mnras, 401, 7
\bibitem[Janiuk \& Czerny(2011)]{Jan11} Janiuk, A. \& Czerny, B.\ 2011, \mnras, 414, 2186
\bibitem[Kallman \& Bautista(2001)]{xstar} Kallman, T. \& Bautista, M.\ 2001, ApJ Suppl., 133, 221
\bibitem[Kara et al.(2016)]{Kara16} Kara, E., Alston, W.~N., Fabian, A.~C., et al.\ 2016, \mnras, 462, 511
\bibitem[Kelly \& Bechtold(2007)]{kelly07} Kelly, B.~C. \& Bechtold, J.\ 2007, \apjs, 168, 1. doi:10.1086/509725
\bibitem[King \& Pounds(2015)]{KP15} King, A., \& Pounds, K. 2015, ARA\&A, 53, 115
\bibitem[Kormendy \& Ho(2013)]{KH13} Kormendy, J., \& Ho, L.~C.\ 2013, \araa, 51, 511
\bibitem[Laha et al.(2014)]{Laha14} Laha, S., Guainazzi, M., Dewangan, G.~C., et al.\ 2014, \mnras, 441, 2613
\bibitem[Laurenti et al.(2020)]{La20} Laurenti, M., Luminari, A., Tombesi, F., et al.\ 2020, \aap in press, arXiv:2011.08212
\bibitem[Longinotti et al.(2015)]{Long15} Longinotti, A.~L., Krongold, Y., Guainazzi, M., et al.\ 2015, \apjl, 813, L39
\bibitem[Luminari et al.(2020)]{L20} Luminari, A., Tombesi, F., Piconcelli, E., et al.\ 2020, \aap, 633, A55
\bibitem[Lusso et al.(2012)]{L12} Lusso, E., Comastri, A., Simmons, B.~D., et al.\ 2012, \mnras, 425, 623
\bibitem[Matthews et al.(2020)]{matthews20} Matthews, J.~H., Knigge, C., Higginbottom, N., et al.\ 2020, \mnras, 492, 5540
\bibitem[Nardini et al.(2015)]{N15} Nardini, E., Reeves, J.~N., Gofford, J., et al.\ 2015, Science, 347, 860.
\bibitem[Nardini \& Zubovas(2018)]{N18} Nardini, E., \& Zubovas, K.\ 2018, \mnras, 478, 2274
\bibitem[Netzer(2013)]{netzer13} Netzer, H.\ 2013, The Physics and Evolution of Active Galactic Nuclei, by Hagai Netzer, Cambridge, UK: Cambridge University Press, 2013
\bibitem[Nomura et al.(2016)]{nomura16} Nomura, M., Ohsuga, K., Takahashi, H.~R., et al.\ 2016, \pasj, 68, 16. doi:10.1093/pasj/psv124
\bibitem[Nomura et al.(2020)]{nomura20} Nomura, M., Ohsuga, K., \& Done, C.\ 2020, \mnras, 494, 3616. doi:10.1093/mnras/staa948
\bibitem[Onori et al.(2017)]{onori17} Onori, F., Ricci, F., La Franca, F., et al.\ 2017, \mnras, 468, L97. doi:10.1093/mnrasl/slx032
\bibitem[Parker et al.(2017)]{P17} Parker, M.~L., Pinto, C., Fabian, A.~C., et al.\ 2017, \nat, 543, 83
\bibitem[Peterson et al.(2004)]{peterson04} Peterson, B.~M., Ferrarese, L., Gilbert, K.~M., et al.\ 2004, \apj, 613, 682. doi:10.1086/423269
\bibitem[Piconcelli et al.(2005)]{P05} Piconcelli, E., Jimenez-Bail{\'o}n, E., Guainazzi, M., et al.\ 2005, \aap, 432, 15
\bibitem[Proga et al.(2000)]{PSK00} Proga, D., Stone, J.~M., \& Kallman, T.~R.\ 2000, \apj, 543, 686
\bibitem[Proga \& Kallman(2004)]{PK04} Proga, D., \& Kallman, T.~R.\ 2004, ApJ, 616, 688
\bibitem[Quera-Bofarull et al.(2020)]{Q20} Quera-Bofarull, A., Done, C., Lacey, C., et al.\ 2020, \mnras, 495, 402
\bibitem[Reeves \& Braito(2019)]{Reeves19} Reeves, J.~N., \& Braito, V.\ 2019, \apj, 884, 80
\bibitem[Reeves et al.(2020)]{Reeves20} Reeves, J.~N., Braito, V., Chartas, G., et al.\ 2020, \apj, 895, 37
\bibitem[Reis \& Miller(2013)]{Reis13} Reis, R.~C., \& Miller, J.~M.\ 2013, \apjl, 769, L7
\bibitem[Reis et al.(2014)]{Reis14} Reis, R.~C., Reynolds, M.~T., Miller, J.~M., et al.\ 2014, \nat, 507, 207
\bibitem[Ricci et al.(2017)]{ricci17m} Ricci, F., La Franca, F., Marconi, A., et al.\ 2017, \mnras, 471, L41. doi:10.1093/mnrasl/slx103
\bibitem[Ricci et al.(2017)]{ricci17} Ricci, F., La Franca, F., Onori, F., et al.\ 2017, \aap, 598, A51. doi:10.1051/0004-6361/201629380
\bibitem[Ripperda et al.(2020)]{Ripp20} Ripperda, B., Bacchini, F., \& Philippov, A.\ 2020, arXiv:2003.04330
\bibitem[Risaliti \& Elvis(2010)]{RE10} Risaliti, G., \& Elvis, M.\ 2010, \aap, 516, A89
\bibitem[R{\'o}{\.z}a{\'n}ska et al.(2014)]{RO14} R{\'o}{\.z}a{\'n}ska, A., Niko{\l}ajuk, M., Czerny, B., et al.\ 2014, New Astronomy, 28, 70
\bibitem[Rybicki \& Lightman(1986)]{RL} Rybicki, G.~B., \& Lightman, A.~P.\ 1986, Radiative Processes in Astrophysics, Chapter 1
\bibitem[Saez \& Chartas(2011)]{SC11} Saez, C., \& Chartas, G.\ 2011, \apj, 737, 91
\bibitem[Saturni et al.(2018)]{Saturni18} Saturni, F.~G., Bischetti, M., Piconcelli, E., et al.\ 2018, \aap, 617, A118
\bibitem[Schurch \& Done(2007)]{schurch07} Schurch, N.~J. \& Done, C.\ 2007, \mnras, 381, 1413. doi:10.1111/j.1365-2966.2007.12336.x
\bibitem[Shakura \& Sunyaev(1973)]{SS73} Shakura, N.~I., \& Sunyaev, R.~A.\ 1973, \aap, 500, 33
\bibitem[Serafinelli et al.(2019)]{Sera19} Serafinelli, R., Tombesi, F., Vagnetti, F., et al.\ 2019, \aap, 627, A121
\bibitem[Sim et al.(2010)]{SI10} Sim, S.~A., Proga, D., Miller, L., et al.\ 2010, \mnras, 408, 1396
\bibitem[Smith et al.(2019)]{Smith19} Smith, R.~N., Tombesi, F., Veilleux, S., et al.\ 2019, \apj, 887, 69
\bibitem[Sturm et al.(2018)]{ST18} Sturm, E., Dexter, J., Pfuhl, O. et al.\ 2018, Nature 563, 657–660
\bibitem[Szanecki et al.(2020)]{Sz20} Szanecki, M., Niedzwiecki, A., Done, C., et al.\ 2020, arXiv e-prints, arXiv:2006.15016
\bibitem[Tilton \& Shull(2013)]{tilton13} Tilton, E.~M. \& Shull, J.~M.\ 2013, \apj, 774, 67. doi:10.1088/0004-637X/774/1/67
\bibitem[Tombesi et al.(2011)]{T11} Tombesi, F., Cappi, M., Reeves, J.~N., et al.\ 2011, \apj, 742, 44
\bibitem[Tombesi et al.(2012)]{T12} Tombesi, F., Cappi, M., Reeves, J.~N., et al.\ 2012, \mnras, 422, L1
\bibitem[Tombesi et al.(2013)]{T13} Tombesi, F., Cappi, M., Reeves, J.~N., et al.\ 2013, \mnras, 430, 1102
\bibitem[Tombesi et al.(2015)]{T15} Tombesi, F., Mel{\'e}ndez, M., Veilleux, S., et al.\ 2015, \nat, 519, 436
\bibitem[Vestergaard \& Peterson(2006)]{vestergaard06} Vestergaard, M. \& Peterson, B.~M.\ 2006, \apj, 641, 689. doi:10.1086/500572
\bibitem[Yuan et al.(2015)]{Yuan15} Yuan, F., Gan, Z., Narayan, R., et al.\ 2015, \apj, 804, 101
\bibitem[Zappacosta et al.(2020)]{Zapp20} Zappacosta, L., Piconcelli, E., Giustini, M., et al.\ 2020, \aap, 635, L5
\end{thebibliography}
\end{document}